\documentclass[aps,
twocolumn,
nofootinbib,
preprintnumbers,
showkeys,
superscriptaddress]{revtex4-2}

\usepackage{graphicx}
\usepackage{dcolumn}
\usepackage{bm}
\usepackage{amsmath}
\usepackage{wasysym}
\usepackage{float}
\usepackage{multirow}
\usepackage{xcolor}
\usepackage{scrextend}
\usepackage[colorlinks=true, 
pdfstartview=FitV, 
bookmarks=true, 
bookmarksnumbered=true, 
breaklinks]{hyperref}
\usepackage{color}
\definecolor{blue}{rgb}{0.0, 0.0, 1.0}
\definecolor{red}{rgb}{1.0, 0.0, 0.0}
\definecolor{royalblue}{rgb}{0.0, 0.14, 0.4}
\hypersetup{linkcolor=royalblue, 
	citecolor=blue, 
	urlcolor=royalblue}

\def\orcid#1{\kern .08em\href{https://orcid.org/#1}{\includegraphics[keepaspectratio,width=0.7em]{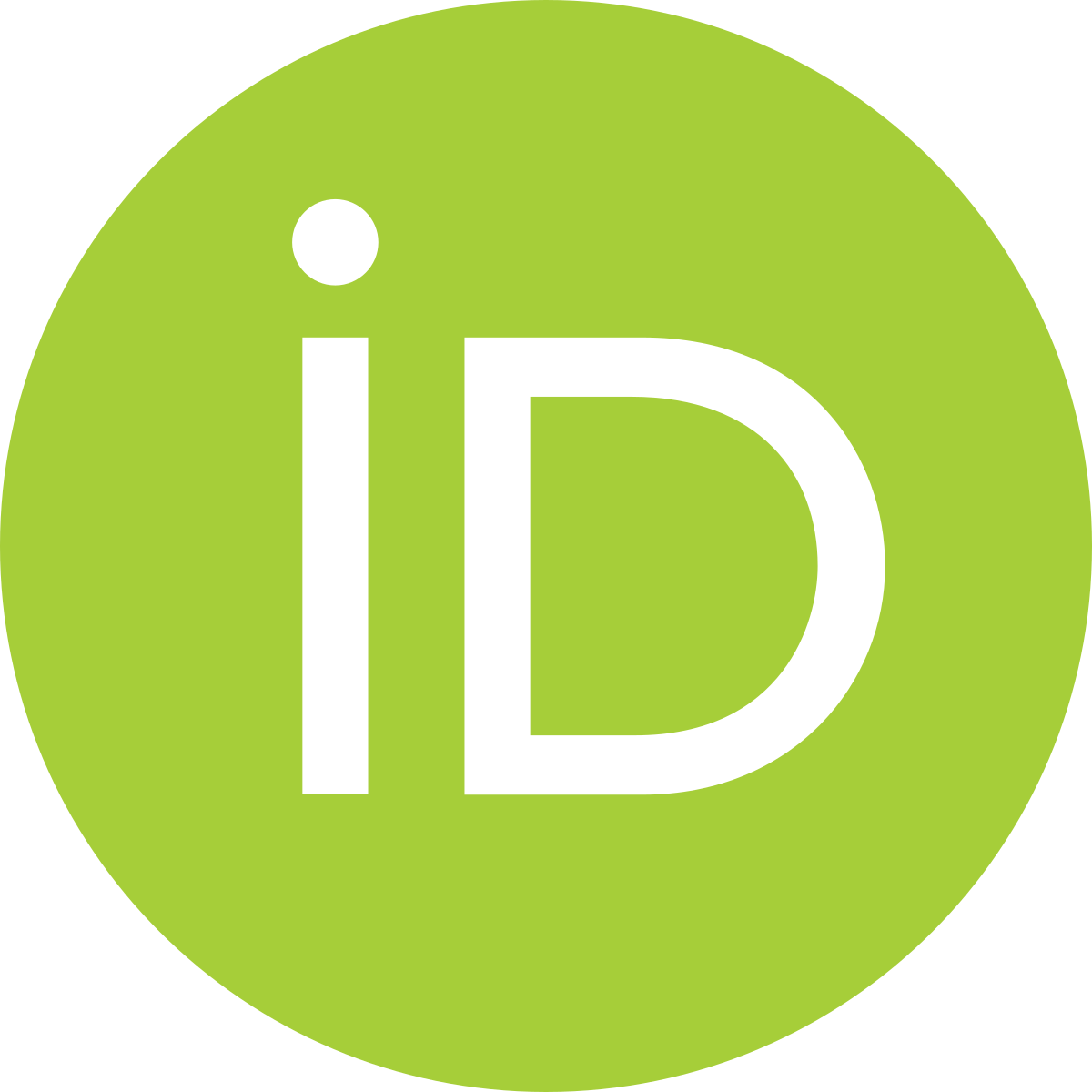}}}

\begin{document}
\title{Formulation of causality-preserving quantum time of arrival theory}

\author{Denny Lane B.
	 Sombillo\orcid{0000-0001-9357-7236}}
\email[]{dbsombillo@up.edu.ph}
\affiliation{National Institute of Physics, University of the Philippines Diliman, Quezon City 1101, Philippines}

\author{Neris I. 
	Sombillo\orcid{0000-0003-1186-6638}}
\affiliation{Department of Physics, School of Science and Engineering, Ateneo de Manila University, \\
Loyolo Heights Quezon City 1108, Philippines}

\date{\today}
\begin{abstract}
	We revisit the quantum correction to the classical time of arrival to address the unphysical instantaneous arrival in the limit of zero initial momentum. In this study, we show that the vanishing of arrival time is due to the contamination of the causality-violating component of the initial wave packet. Motivated by this observation, we propose to update the temporal collapse mechanism in [Galapon E. A. 2009, Proc. R. Soc. A.46571–86] to incorporate the removal of causality-violating spectra of the arrival time operator. We found that the quantum correction to the classical arrival time is still observed. Thus, our analysis validates that the correction is an inherent consequence of quantizing a time observable and is not just some mathematical artifact of the theory. We also discuss the possible application of the theory in describing point interactions in particle physics and provide a possible explanation to the observed neutron's lifetime anomaly. 
\end{abstract}		

\maketitle

\section{Introduction}
The quantum nature of time is one of the unresolved and controversial topics in the foundation of quantum mechanics. Not to mention that the quantum time problem is as old as quantum mechanics itself \cite{Muga2008}. One major concern is whether to treat time as a parameter or as an observable. The latter treatment captures the interest of several studies, which are motivated by different experimental developments and proposals of new experiments that can be used to test different theories \cite{Reflection,Htunneling,Eckle,Sombillo2018,PablicoTunnel,Flores:2022pjn,DasArrival,MacconePRA,Maccone2020,ExpTOA1}. 

Perhaps the most popular topic in the quantum time problem is the measurement of time of arrival. The problem can  be unambiguously formulated because we can write down the classical expression for the arrival time. One can then apply some quantization rule to construct a quantum time of arrival (QTOA) operator and study its spectral properties. Probability distributions and expectation values can be calculated and compared with experiment if there is any.

Once we have constructed our QTOA operator, 
a natural direction of inquiry is to find any quantifiable differences with the classical counterpart. Some studies are done in this area where the quantum corrections are calculated for different kinds of systems and different quantization schemes \cite{Flores:2018ken,PhysRevA.94.032123,Flores:2022vvk,Pablico:2022vjq}. For the case of free particle, one can calculate the expectation value of a Weyl quantized free time of arrival expression. It was shown in \cite{Galapon2009} that, for a Gaussian wave packet, the quantum correction factor is a function of the dimensionless quantity $k_0\sigma_0$ where $k_0$ is the incident wave number and $\sigma_0^2$ is the initial position variance. However, it was found that the quantum correction factor vanishes as $k_0$ approaches zero. The vanishing of arrival time at zero incident momentum is already pointed out in \cite{DasNoth2021}, suggesting that such result precludes the use of usual operator treatment of time observable.  
In this work, we show that the vanishing of the quantum arrival time can be removed by applying the causality requirement to the temporal collapse mechanism in \cite{Galapon2008.0278}. One of the important results of this study is that, even after imposing causality, the quantum correction to the classical time of arrival is still present. We then discuss the possibility of explaining the neutron lifetime anomaly as a consequence of quantum correction to time observable.

\section{Quantum Time of Arrival}
In this section, we quickly summarize the QTOA theory for a free particle. The classical expression for the arrival time at point $X$ of a particle with mass $\mu$, momentum $p$, and initial position $q$ is given by $\tau_{class}=\mu(X-q)/p$. Using symmetric quantization, the corresponding QTOA operator is obtained \cite{AharonovBohm1961}
\begin{equation}
	\hat{T}=\dfrac{\mu}{2}\left[(X-\hat{q})\hat{p}^{-1}+\hat{p}^{-1}(X-\hat{q})\right].
	\label{eq:freeTOAoperator}
\end{equation}
It can be shown that the operator $\hat{T}$ satisfies the expected commutation relation $[\hat{T},\hat{H}]=-i\hbar$ with the free Hamiltonian $\hat{H}=\hat{p}^2/2\mu$ \cite{AharonovBohm1961,Giannitrapani1997,Galapon2004}. 

\begin{figure*}[ht!]
	\includegraphics[width=0.246\textwidth]{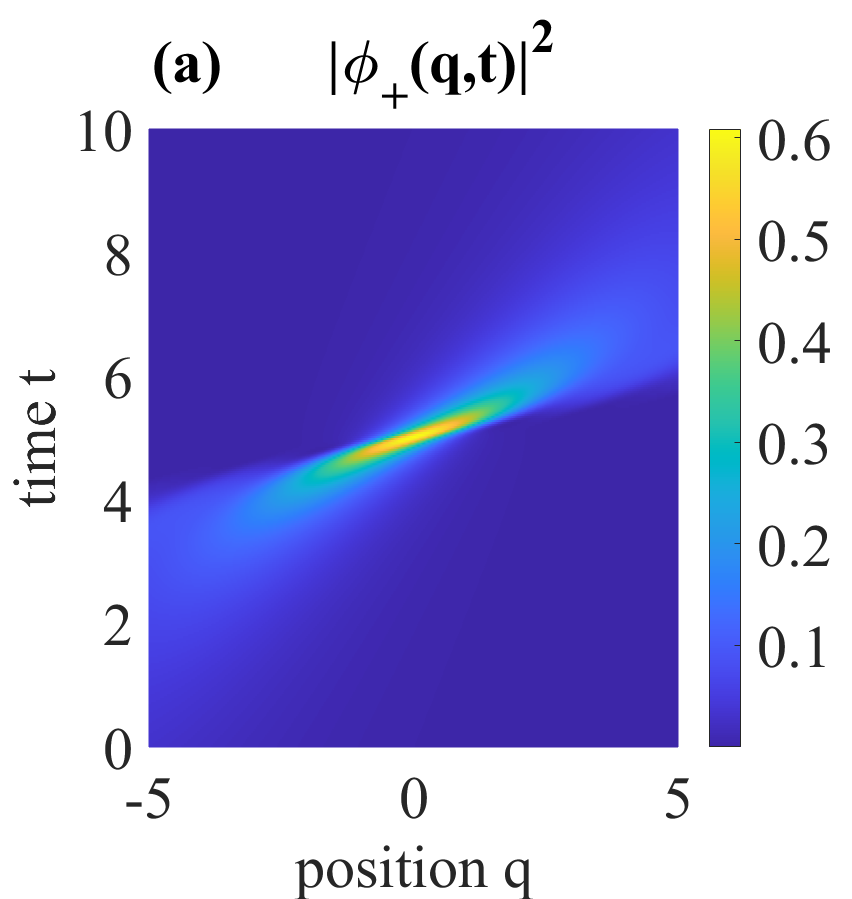}
	\includegraphics[width=0.246\textwidth]{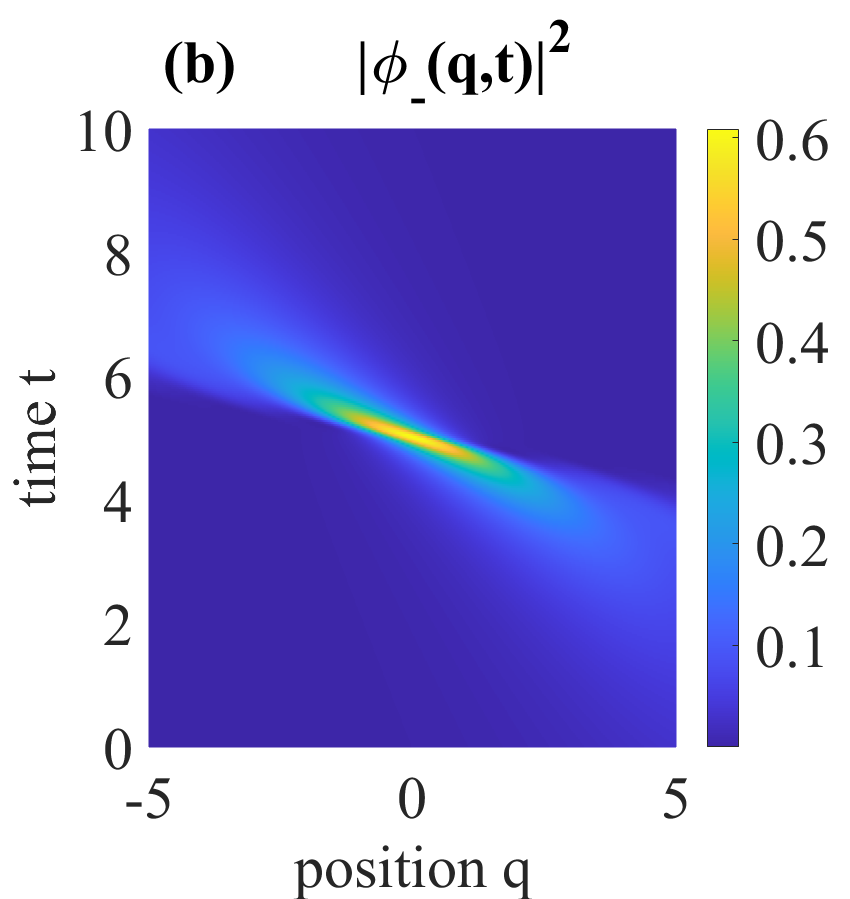}	
	\includegraphics[width=0.246\textwidth]{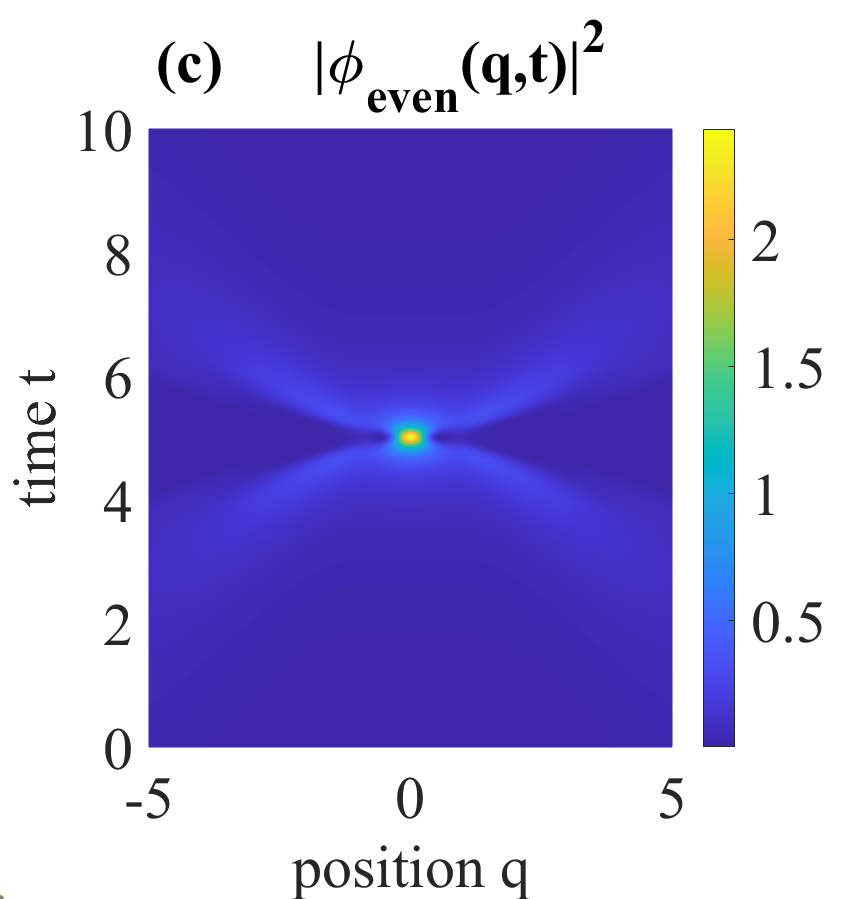}
	\includegraphics[width=0.246\textwidth]{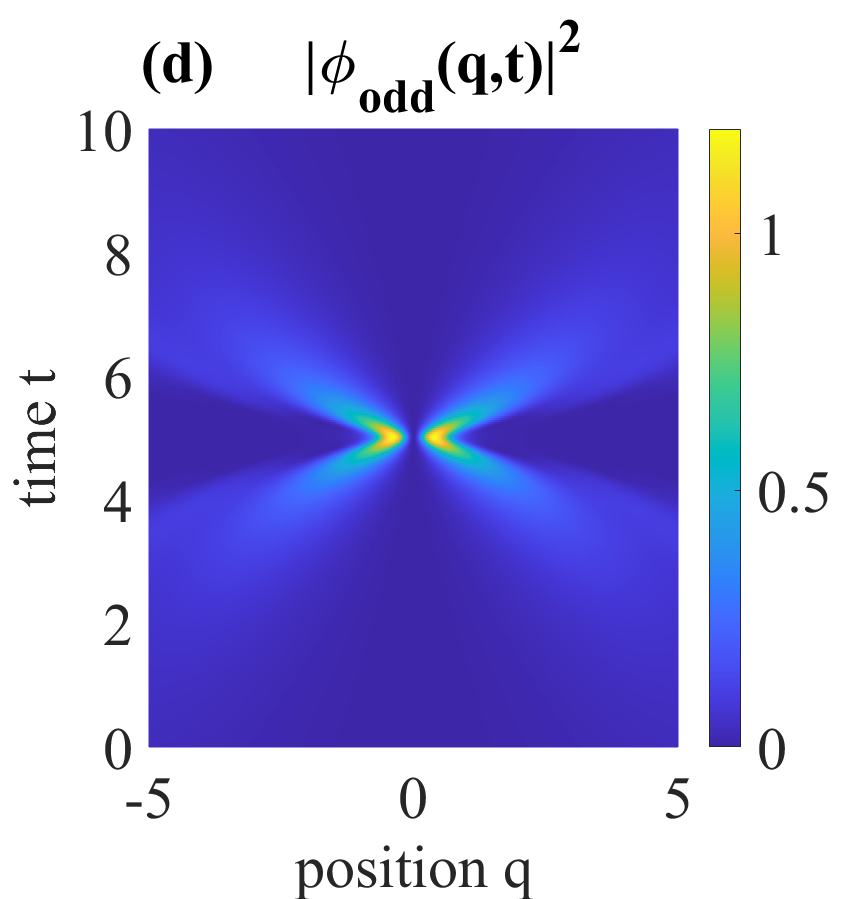}	
	\caption{The regulated free evolution of the TOA operator eigenfunctions for the left-right basis in (a) and (b), and the even-odd in (c) and (d). For all cases we set $X=0$ and $\tau=5.0$. All the other constants are set to unity.}
	\label{fig:unitarycollapse}
\end{figure*}

Upon removing the $p=0$ in the momentum spectrum, the degenerate eigenfunctions of $\hat{T}$ with eigenvalue $\tau$ can be expressed in terms of the right-left basis \cite{MugaLeavensPalao1998}
\begin{equation}
	\phi_{\pm}(p)=\left\langle p|\tau_\pm\right\rangle
	=\sqrt{\dfrac{|p|}{\mu}}\dfrac{e^{-iXp/\hbar}}{\sqrt{2\pi\hbar}}e^{i\tau p^2/2\mu\hbar}\Theta(\pm p)
	\label{eq:pmbasis}
\end{equation}
or the even-odd basis \cite{SOMBILLO2016261}
\begin{equation}
	\begin{split}
		\phi_{\text{even}}(p)&=\left\langle p|\tau_\text{even}\right\rangle
		=\sqrt{\dfrac{|p|}{\mu}}\dfrac{e^{-iXp/\hbar}}{\sqrt{2\pi\hbar}}e^{i\tau p^2/2\mu\hbar} \\
		\phi_{\text{odd}}(p)&=\left\langle p|\tau_\text{odd}\right\rangle
		=\sqrt{\dfrac{|p|}{\mu}}\dfrac{e^{-iXp/\hbar}}{\sqrt{2\pi\hbar}}e^{i\tau p^2/2\mu\hbar}\text{sgn}(p). 	
	\end{split}
	\label{eq:oddevenbasis}
\end{equation}
The $\Theta(\pm p)$ and $\text{sgn}(p)$ are the Heaviside and the signum functions, respectively. Both bases evolve via the free unitary evolution operator $\hat{U}(t)=e^{-it\hat{p}^2/2\mu\hbar}$, smoothly transitioning from a spread out distribution into a localized wave function at the point $q=X$ at the eigenvalue time $t=\tau$ as shown in Fig.~\ref{fig:unitarycollapse}. We refer to this behavior as unitary collapse. The distributions at $t=\tau$ for Fig.~\ref{fig:unitarycollapse}(a)-(c) with singular support at $q=X$ (a $\delta(q-X)$-like distribution) fits the description of particle appearance. Notice that when the particle's path is known, the unitary collapse always result into particle appearance at the arrival point. This is not the case when the path is not known, i.e., the unitary collapse may result into particle detection as shown in Fig.~\ref{fig:unitarycollapse}(c) or something else as shown in Fig.~\ref{fig:unitarycollapse}(d). The position distribution at $t=\tau$ of Fig.~\ref{fig:unitarycollapse}(d) fits the description of non-detection at the arrival point \cite{SOMBILLO2016261}. Specifically, we have a non trivial distribution (we know the particle is somewhere) that is zero at the arrival point (the detector will not click) and with a vanishing position variance (no position uncertainty). Thus, the quantization of the classical arrival time expression gives us something more than just particle appearance at the arrival point.

The natural unitary evolution from a spread out wave function into a distribution with singular support changes the ordering of collapse and free evolution in a time of arrival measurement. In the conventional description, the initial wave function evolves unitarily and then collapses at the instant of detection. In \cite{Galapon2008.0278}, the ordering is reversed. The initial wave function collapses into one of the QTOA operator eigenfunctions at the instant of preparation (or particle creation) and then the eigenfunction naturally evolves into a function with singular support at the position of the detector. We call this the Galapon collapse mechanism (GCM) \cite{Footnote1}. The GCM is a more natural description since the time of preparation is the only instance when an external agent is altering the system. After the preparation, there is no reason for a random collapse to happen just to have a detection. That is, the observed detection should be a direct result of unitary evolution. For this matter, the theory of QTOA operator with GCM provides a reasonable description of the time of arrival measurement.

\section{Vanishing arrival time}
The demonstration that there exists quantum correction to the classical time of arrival can be facilitated by comparing the expectation value of QTOA operator with the classical value. The likelihood that an initial state $|\psi_0\rangle$ will collapse to one of the QTOA eigenstate $|\tau\rangle$ is given by $|\langle\tau|\psi_0\rangle|^2$. It follows that the measured time duration can be expressed as an expectation value of $\hat{T}$ with respect to the initial state $|\psi_0\rangle$. In position representation, we have
\begin{equation}
	\begin{split}
		\tau_{\text{quant}}&=
		\langle\psi_0|\hat{T}|\psi_0\rangle \\
		&=\int_{-\infty}^{+\infty}dq
		\psi^*_0(q)
		\int_{-\infty}^{+\infty}dq'
		\langle q|\hat{T}|q'\rangle
		\psi_0(q').
	\end{split}
	\label{eq:tauquant}
\end{equation}
From \eqref{eq:freeTOAoperator}, the kernel $\langle q|\hat{T}|q'\rangle$ requires the evaluation of $\langle q|\hat{p}^{-1}|q'\rangle$. One can perform the Cauchy principal value integration of $\langle q|\hat{p}^{-1}|q'\rangle$ since $p=0$ is removed from the momentum spectrum. The result is $\langle q|\hat{T}|q'\rangle=i\mu\left(2X-q-q'\right)\text{sgn}(q-q')/4\hbar$. 

It is more convenient to separate the momentum dependence in the initial state. We let $\psi_0(q)=\varphi(q)e^{ik_0q}$ where $\hbar k_0$ is the initial momentum and choose a $\varphi(q)$ such that the centroid is at $q=q_0$ with variance $\sigma_0^2$. To ensure that the momentum dependence is only in the factor $e^{ik_0q}$, we impose $\int dq\varphi_0(q)^*\varphi_0'(q)=0$. It is worth noting that the chosen initial state can only collapse into the right-left basis in \eqref{eq:pmbasis} with the accompanying unitary evolution in Fig.~\ref{fig:unitarycollapse}(a)-(b). Now, the integral in \eqref{eq:tauquant} can be simplified by introducing the following dimensionless variables: $u=\left[(q+q)'/2-q_0\right]/\sigma_0$ and $v=k_0(q-q')$. Transforming the $qq'$ integral into $uv$, we obtain the Jacobian $\sigma_0/k_0$. The $k_0$ of the Jacobian can be combined with the $\hbar$ in the kernel $\langle q|\hat{T}|q'\rangle$ giving the momentum $p_0=\hbar k_0$. Except for the $\sigma_0$, all the other constants combine to form the classical arrival time $\tau_\text{class}=\mu(X-q_0)/p_0$. For a Gaussian wave packet $\varphi(q)=(\sqrt{2\pi}\sigma_0)^{-1/2}
e^{-(q-q_0)^2/4\sigma_0^2}$, the $u$ and $v$ integrals factorize with the $u$ integral canceling the $\sigma_0$ of the Jacobian giving us
\begin{equation}
	\tau_{\text{quant}}=
	\tau_{\text{class}}
	\int_{0}^{\infty}dv\sin v
	\exp\left(-\dfrac{v^2}{8k_0^2\sigma_0^2}\right).
	\label{eq:gaussiancorrection}
\end{equation} 
For any non-zero $k_0\sigma_0$, the $\tau_{\text{quant}}$ gives a non-zero measurable value. The integral in \eqref{eq:gaussiancorrection}, which is also the ratio $\tau_{\text{quant}}/\tau_{\text{class}}$, is referred to as the quantum correction factor. One noticeable feature of the correction factor is that it is dependent only on $k_0\sigma_0$ or, equivalently, on the information of initial state.

\begin{figure}[ht!]
	\includegraphics[width=0.40\textwidth]
	{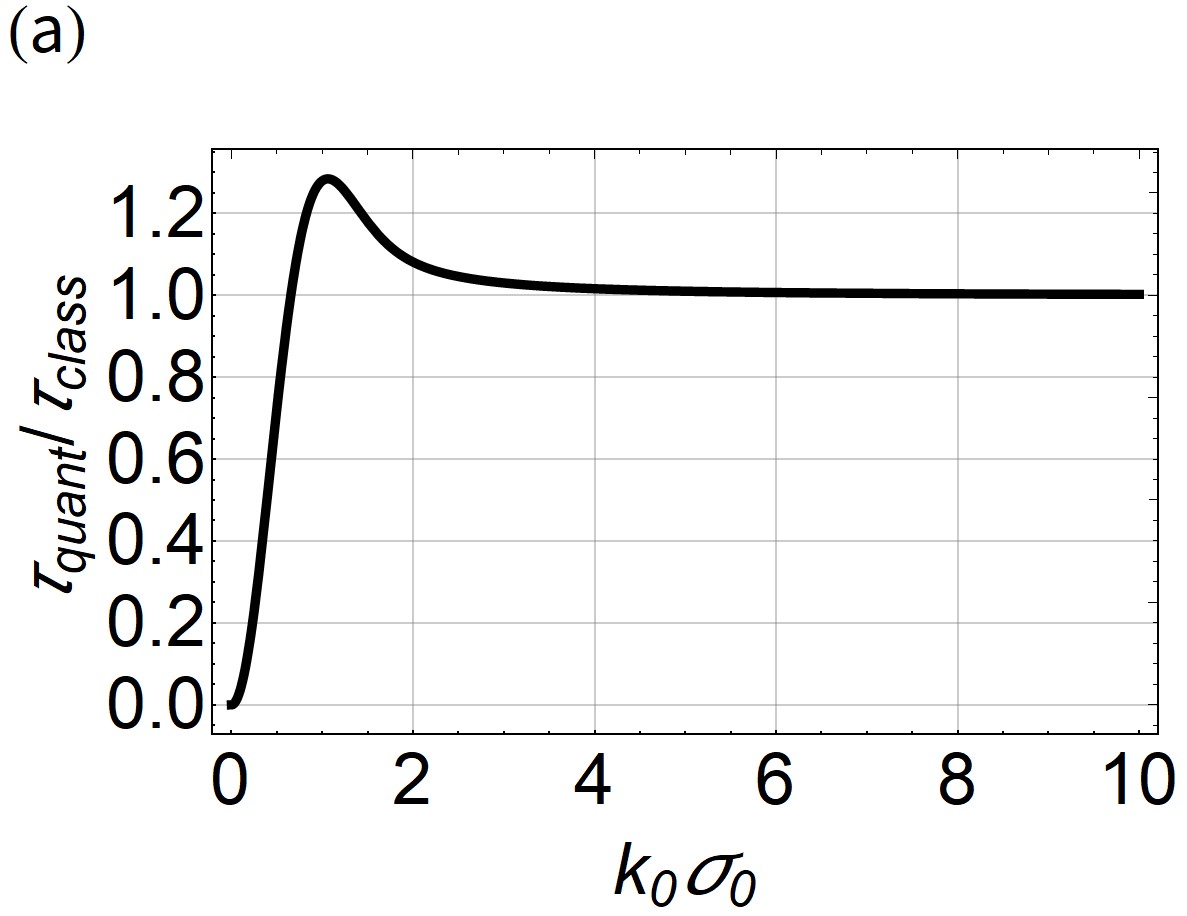}
	\includegraphics[width=0.40\textwidth]
	{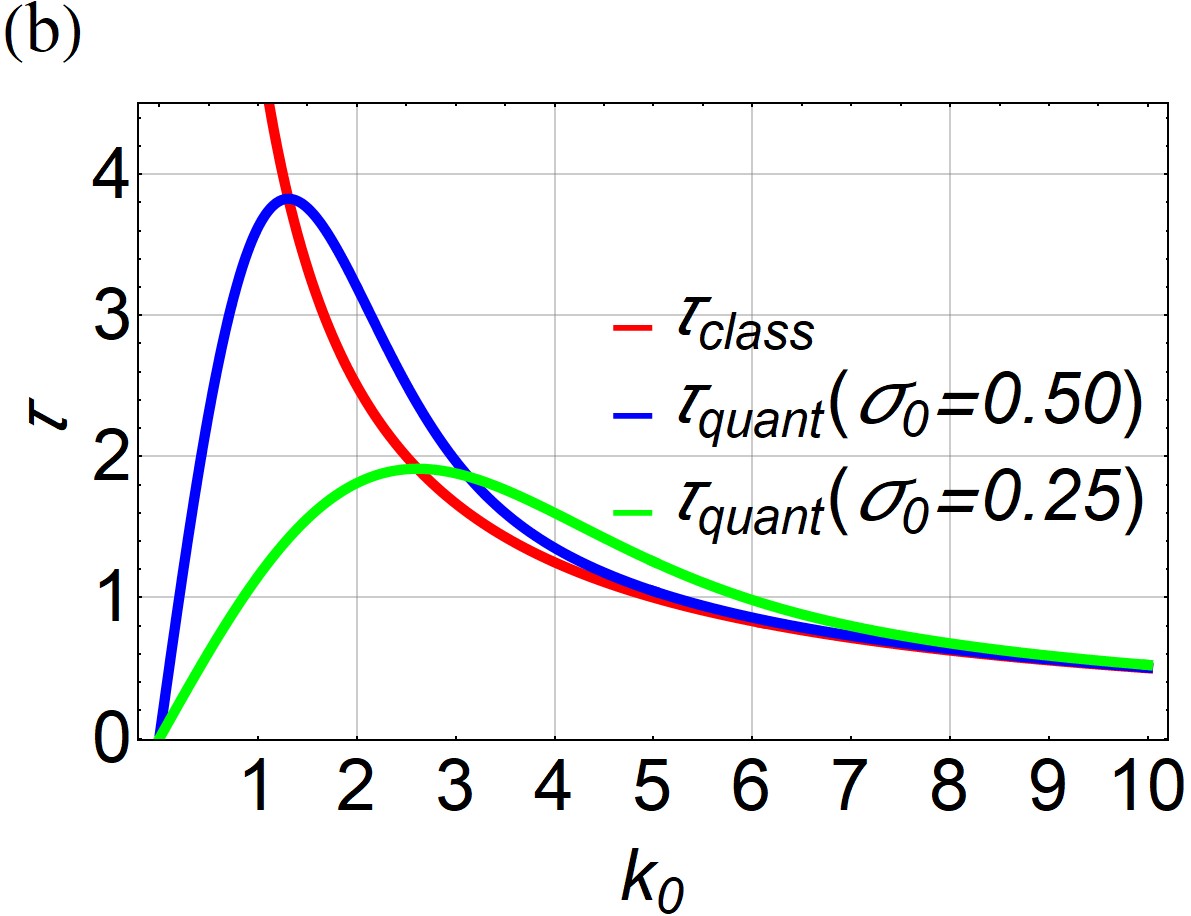}
	\caption{(a) The quantum correction integral factor in Eq.~\eqref{eq:gaussiancorrection} as a function of $k_0\sigma_0$. 
		(b) Quantum arrival times for different Gaussian wave packets with different position variances $\sigma_{0}^2$. 
		For all cases, we set $q_0=-5$ and the arrival point at $X=0$. The classical arrival time is shown for comparison.}
	\label{fig:correction}
\end{figure}

Figure~\ref{fig:correction}(a) shows the behavior of the quantum correction factor as a function of the dimensionless quantity $k_0\sigma_0$. The high $k_0\sigma_0$ region correspond to the classical limit where the de Broglie wavelength, $2\pi/k_0$, is too small compared to $\sigma_0$. The quantum correction is prominent (far from 1) in the region where the low energy de Broglie wavelength is comparable to the size of the wave packet, i.e. $\sigma_0\sim 1/k_0$. However, at some point, the correction begins to drop and approach zero. Surprisingly, recent work involving relativistic treatment still gives vanishing correction factor in the limit of zero initial momentum \cite{Flores:2022vvk}. One can argue that the vanishing of the correction factor is caused by $\tau_{\text{class}}\rightarrow \infty$ as $\hbar k_0\rightarrow 0$ and that $\tau_{\text{quant}}$ will give a non-vanishing result. This is not the case as shown in Fig.~\ref{fig:correction}(b) where $\tau_{\text{quant}}$ vanishes as $k_0\rightarrow 0$ suggesting instantaneous arrival. It was already pointed out in \cite{Galapon2009} that the vanishing of $\tau_{\text{quant}}/\tau_{\text{class}}$ is due to the negative momentum component of the initial wave packet but no further justification was made whether the behavior is physical or not. Consequently, it is also imperative to discuss whether the calculated correction is a consequence of time quantization or just a mathematical artifact associated with negative momentum component. Additionally, a physical argument must be added to justify the removal of the negative momentum component. It turns out that the inclusion of causality in the GCM can provide a coherent picture.

\section{The Role of Causality}
Causality plays a very special role in any physical theories. Violating causality will give wrong commutation relations for the boson and fermion operators which will result into wrong statistics \cite{Peskin,PauliSpin,PCT}. Causality is also one of the guiding principles in constraining the analytic structure of S-matrix in order to interpret and parametrize scattering data \cite{Eden:1966dnq,Newton:1982qc,Sombillo:2021ifs}. In the formulation of a theory involving time, causality should play a central role. 

The GCM provides an operational interpretation for $t=0$ in the QTOA theory. Specifically, $t=0$ is aptly interpreted as the instant of preparation where the initial wave function collapses into one of the QTOA operator eigenfunctions. Knowing that the QTOA eigenfunction can evolve into a function representing particle detection at time $t=\tau$, then the only causally permitted scenario is when $\tau>0$. That is, causality requires that particle detection cannot precede state preparation or particle creation. For an initial wave packet considered in this work, the negative momentum components correspond to negative arrival times since the spatial support of the Gaussian is located left of the arrival point. Thus, we need to remove the component of the momentum that violates causality.  

The removal of the causality-violating term of the initial state is more transparent in the momentum representation. Equation~\eqref{eq:tauquant} is now
\begin{equation}
	\tau_{\text{quant}}=\int_{-\infty}^{+\infty}dp\langle\psi_0|p\rangle
	\langle p|\hat{T}|\psi_0\rangle
	\label{eq:freeTOAmomentum}
\end{equation}
where
\begin{equation}
	\langle p|\hat{T}|\psi_0\rangle
	=-i\mu\hbar\left(\dfrac{1}{p}\dfrac{d}{dp}
	-\dfrac{1}{p^2}+i\dfrac{X}{\hbar p}\right)\tilde{\psi}_0(p).
	\label{eq:TOAdiffoprtr}
\end{equation}
We use the same initial wave packet but in its momentum representation:
\begin{equation}
	\begin{split}
	\langle p|\psi_0 \rangle &=\tilde{\psi}_0(p) \\
	&=\left(\dfrac{2\sigma_0^2}{\pi\hbar^2}\right)^{1/4}
	e^{-(p-p_0)^2\sigma_0^2-i{(p-p_0)q_0}/{\hbar}}.
	\end{split}
	\label{eq:psi0_p_rep}
\end{equation}
The parameters appearing in \eqref{eq:psi0_p_rep} are the same parameters in the position representation of our chosen initial state.
After differentiating, we obtain the following explicit form
\begin{equation}
	\begin{split}
	 \tau_{\text{quant}} &= -i\mu\hbar\int_{-\infty}^{+\infty}\dfrac{dp}{p^2} \\
	 \times&|\tilde{\psi}_0(p)|^2
	\left[
	{-2p(p-p_0)\sigma_0^2+\dfrac{ip(X-q_0)}{\hbar}-1}
	\right].
	\end{split}
	\label{eq:qtoa_p_rep}
\end{equation}

Certainly, the integrand is ill-defined at $p=0$. And even if we removed $p=0$ in the momentum spectrum, the integral will still diverge when the peak of the momentum Gaussian distribution is close to zero. For this reason, we introduce a causality-enforcing regulator of the form $\exp(-\epsilon/p)$ where $\epsilon>0$. The chosen regulator does not affect the large positive momentum component since $\exp(-\epsilon/p)\rightarrow 1$ as $p\rightarrow+\infty$. That is, we expect that the classical limit (small de Broglie wavelength region) is not contaminated by the regulator. Furthermore, the regulator ensures that the negative momentum integration is still divergent, mimicking the expectation that if the particle is moving away from the detector, then the waiting time for detection is infinity. 

\begin{figure}[ht!]
	\includegraphics[width=0.45\textwidth]
	{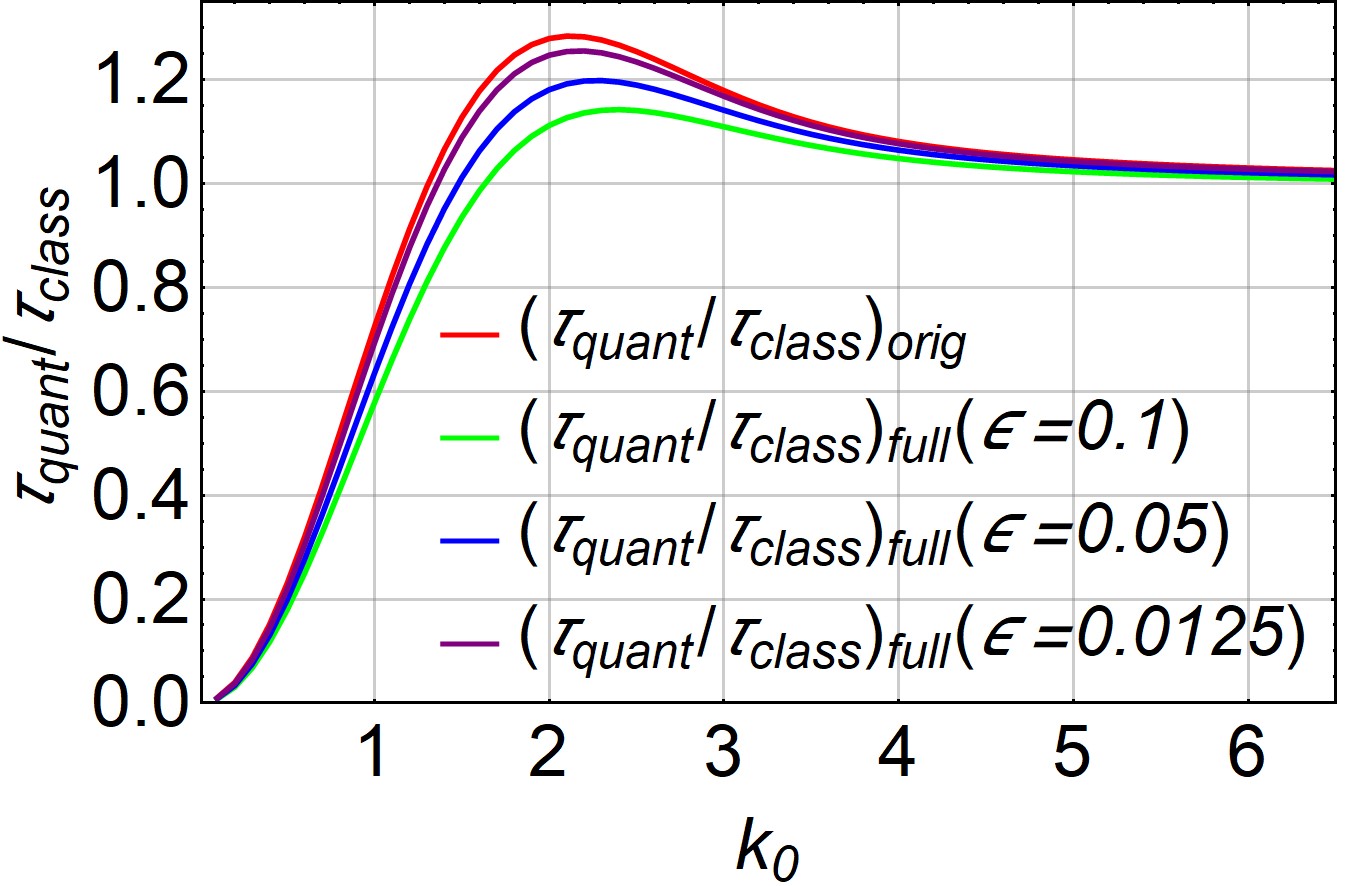}
	\caption{The full integral in \eqref{eq:qtoa_p_rep} with the relaxed regulator $e^{-\epsilon/|p|}$ approaching the original behavior of the quantum correction in \eqref{eq:gaussiancorrection} as $\epsilon\rightarrow 0$. The position variance $\sigma_0^2=(0.5)^2$ is used for all cases.}
	\label{fig:correction2}
\end{figure}

As a sanity check, we verify first if the original result in \eqref{eq:gaussiancorrection} would be recovered using the regulated integral in \eqref{eq:qtoa_p_rep}. For this part, we temporarily relaxed the regulator to $\exp(-\epsilon/|p|)$ just to extract some finite value from the negative momentum integration. Figure~\ref{fig:correction2} shows that the two correction factors agree as we let $\epsilon\rightarrow 0$. The result also demonstrates the equivalence of expectation values calculated in the position and in the momentum representations.
For the subsequent analyses, setting $\epsilon$ to some small value will suffice to see the behavior of causality-preserving $\tau_{\text{quant}}$. 
\begin{figure}[ht!]
	\includegraphics[width=0.45\textwidth]
	{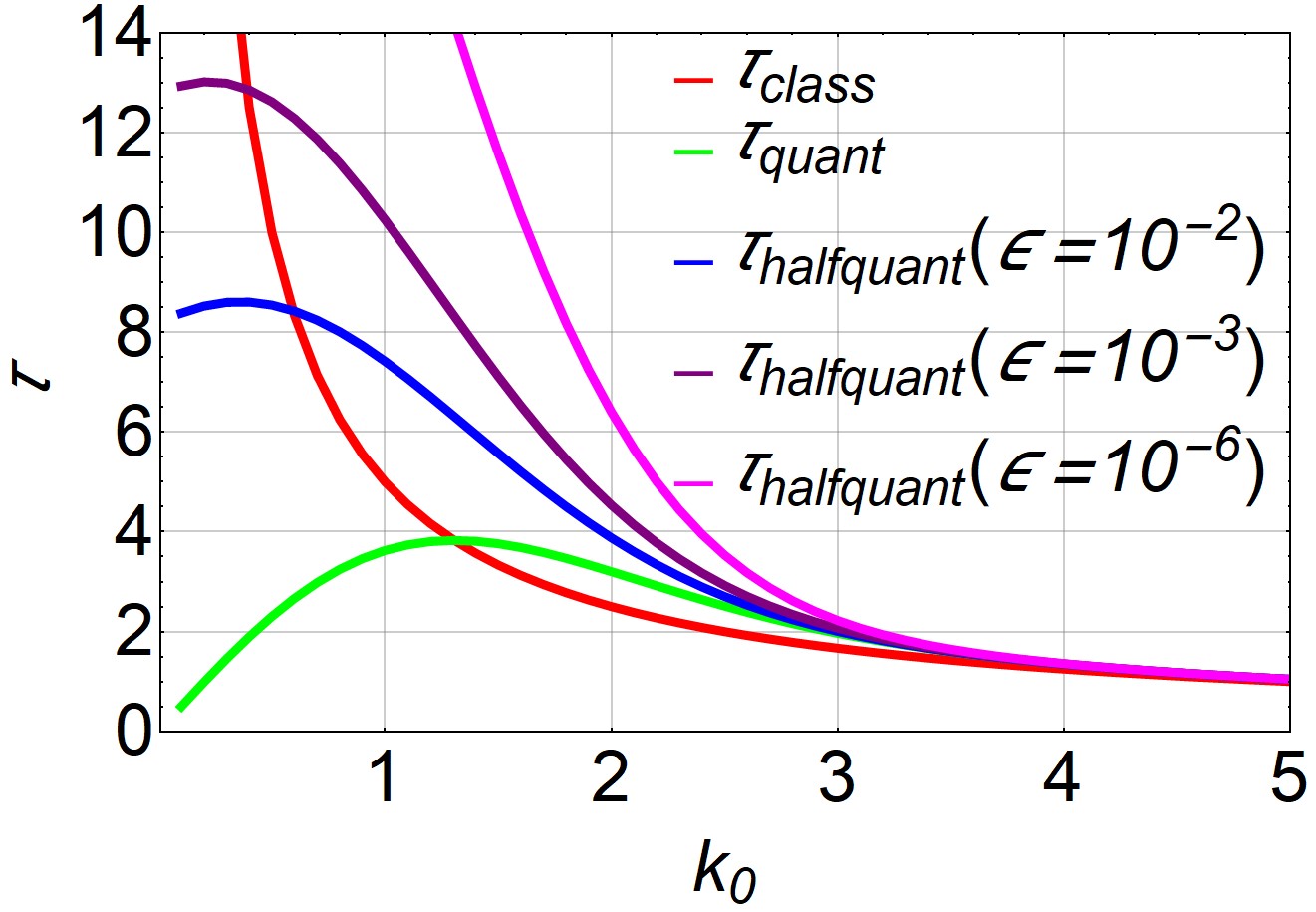}
	\caption{The causality-preserving quantum arrival time with regulator $e^{-\epsilon/p}$ approaching unity. The classical and the causality-violating quantum arrival times are shown for comparison. The position variance $\sigma_0^2=(0.5)^2$ is used for all the quantum cases.}
	\label{fig:correction3}
\end{figure}

We can now check if the quantum correction would remain after the removal of causality-violating component of the initial state. 
This is done by replacing the lower limit of integration in \eqref{eq:qtoa_p_rep} with $0$. The result is shown in Figure~\ref{fig:correction3}. 
Some important observations are worth mentioning.
First, for all values of $\epsilon$, the results all agree with the classical arrival time in the classical limit. This should be expected since the contribution of the negative momentum component is minimized as the peak of the Gaussian momentum distribution moves away from $k_0=0$.
Second, the nonphysical instantaneous arrival at $k_0=0$ is no longer present for the causality-preserving $\tau_{\text{quant}}$. Notice also that as the regulator is removed, the vertical intercept approaches infinity similar to the expected behavior of classical arrival time. Lastly, despite the similarity with classical arrival time as $k_0\rightarrow 0$ the quantum correction is still observed but the original result in \eqref{eq:gaussiancorrection} underestimated the correction. Thus, we can say that there is a range of incident initial momentum where quantum correction to the classical arrival time can be measured.

 \begin{figure}[ht!]
	\includegraphics[width=0.45\textwidth]
	{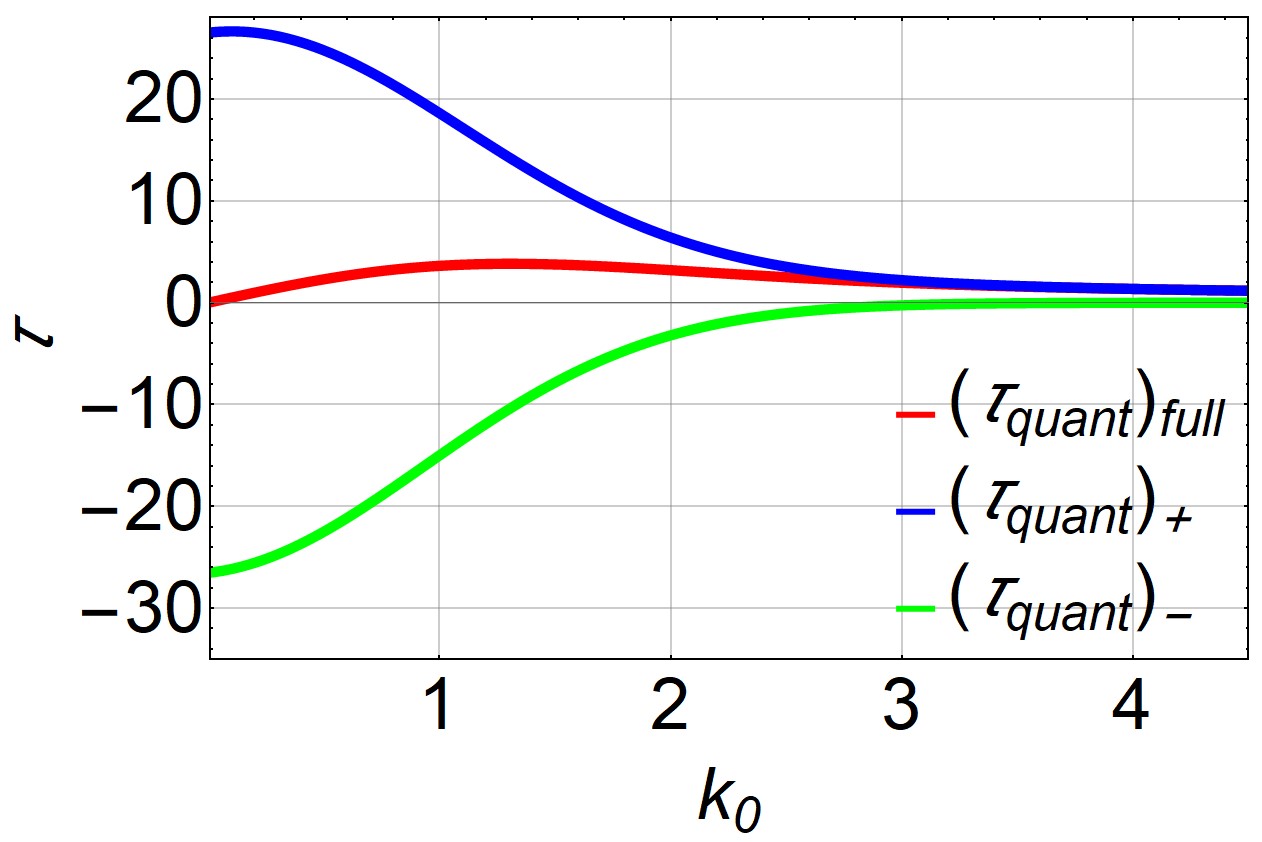}
	\caption{The causality-preserving and causality-violating arrival times. The combined result is also shown for comparison.}
	\label{fig:negativep}
\end{figure}

We also verify if the negative momenta were indeed causality-violating ($\tau<0$) and check if there were any regions where the negative momentum contributes to positive arrival time. We relaxed once again our regulator into $\exp(-\epsilon/|p|)$ and evaluate the other half of the integral in \eqref{eq:qtoa_p_rep}. Figure~\ref{fig:negativep} shows the arrival time associated with negative momentum. It is clear that, for the system considered so far, the regulated causality-violating part of the momentum distribution gives rise to non-positive arrival time. Observe also that the two regulated $\tau_{\text{quant}}$ cancels out at $k_0=0$, resulting to instantaneous arrival if one did not remove the causality-violating component. Also, as the causality-preserving  $\tau_{\text{quant}}$ approaches the classical limit, the causality-violating part vanishes. Thus, no physical features are removed from the causality-preserving quantum arrival time when we drop the negative momentum part of the initial wave packet.

It still remains to ask on what is the meaning of negative arrival time.
In \cite{DasNoth2021}, it was proposed that the distribution associated with the negative arrival time be interpreted as non-detection probability. For this proposal to be acceptable, the QTOA eigenfunction should not unitarily reduce to a Dirac-delta like distribution for all possible values of parametric time $t$. However, if we set the eigenvalue time to some negative values, $\tau<0$, we still observe similar behavior as shown in Fig.~\ref{fig:unitarycollapse} but with peak appearing on the negative parametric time $t=\tau<0$. This means that the negative arrival time still gives rise to particle appearing at $q=X$. Now, with $t=0$ as the preparation time, the negative arrival time is not physically meaningful because it implies violation of causality. This argument is in direct contrast to what is being proposed in \cite{DasNoth2021} because non-detection is a physical outcome of an arrival time measurement. Thus, the only plausible explanation for negative arrival time is that it is an unphysical mathematical feature of the theory, which can be removed by imposing causality. For a more detailed discussion of non-detection in a QTOA measurement, we direct the reader to Ref.\cite{SOMBILLO2016261}.

Care must be taken when extracting the quantum arrival time for an initial wave packet with two different momentum peaks. For this case, one needs to refer to the evolution of  the even eigenfunctions in Fig.~\ref{fig:unitarycollapse}. We no longer think in terms of positive or negative momentum as the causality-violating part of the initial state. It is now the outgoing flux relative to the arrival point that will give rise to unphysical negative arrival time. At the moment, removal of causally-violating component for the mentioned case is not yet considered and will be addressed elsewhere.

 \begin{figure}[ht!]
	\includegraphics[width=0.45\textwidth]
	{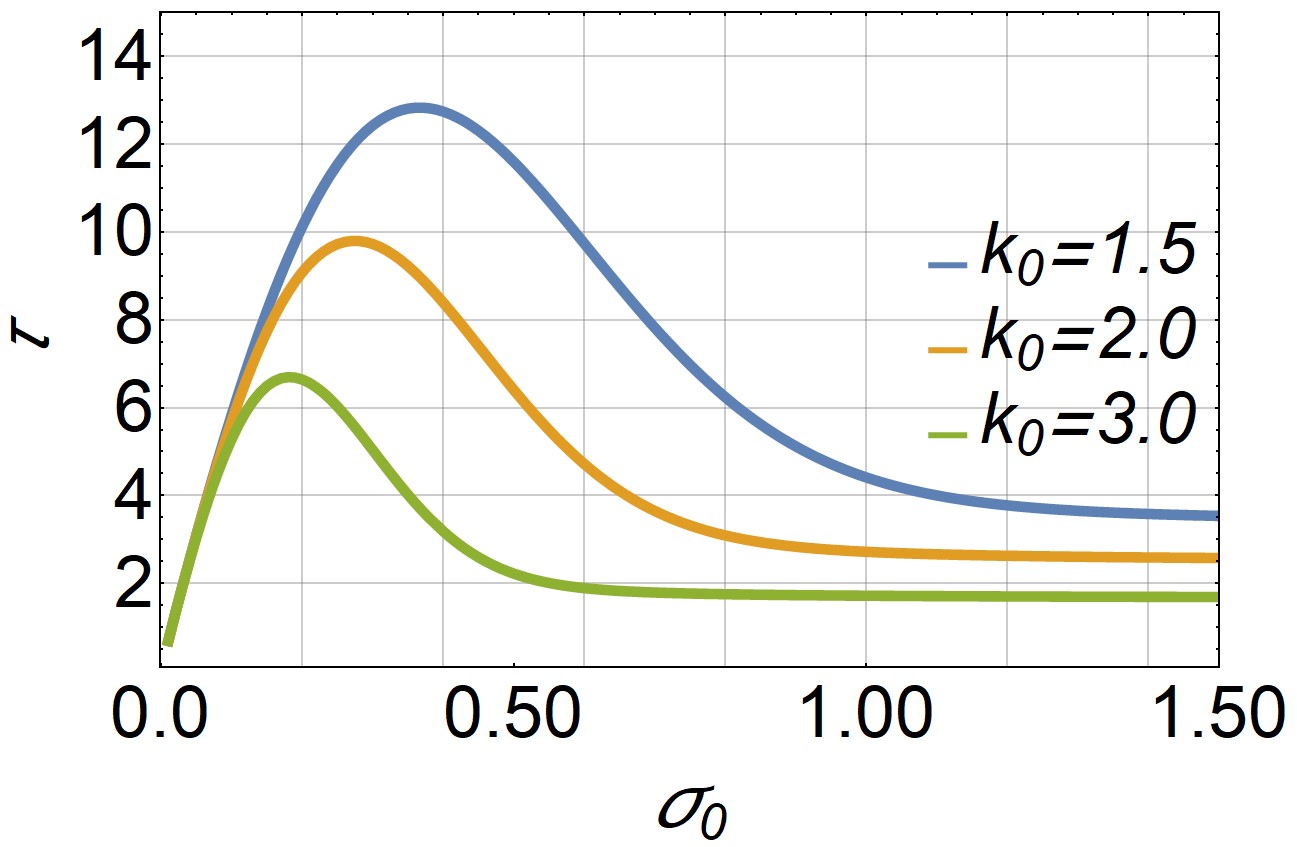}
	\caption{The causality-preserving quantum arrival time as a function of initial position variance. The values approached by the plots for large $\sigma_0$ correspond to the classical arrival times.}
	\label{fig:sigmaTOA}
\end{figure}

Recall that the quantum correction depends on two quantities associated with the initial state. One is the incident momentum $\hbar k_0$, which we have thoroughly discussed, and the other is the position variance $\sigma_0^2$. We have systematically removed the unphysical instantaneous arrival when $k_0=0$ but $\sigma_0\rightarrow0$ will still give rise to vanishing arrival time. Figure~\ref{fig:sigmaTOA} shows the causality-preserving QTOA for different incident momenta. The vanishing of arrival time here is no longer associated to causality but a consequence of the uncertainty principle and the lack of a speed upper bound in quantum mechanics. Specifically, as we let the position variance vanish, the width of the momentum distribution gets larger. The particle will have momentum component that can exceed the speed of light resulting to vanishing arrival time. It is possible that the turning points seen in Figure~\ref{fig:sigmaTOA} are the values of $\sigma_0$ where the large momentum component starts to dominate. In any case, before the turning point, there is a noticeable correction to the classical arrival time as $\sigma_0$ gets smaller.
This implies that the conclusion of \cite{Galapon2009} still holds. That is, different sizes of wave packets result into different arrival times despite having the same incident momentum. Thus, combining all our observations, the causality-preserving quantum correction to arrival time is dependent on the initial preparation of the system. 

\section{Possible Application}
The QTOA theory composed of GCM, unitary collapse in Fig.~\ref{fig:unitarycollapse}, and the recent proposal of imposing causality goes beyond the time of arrival measurement. To illustrate this, consider a simple three-point interaction vertex of some scalar field theory where the interaction Lagrangian is $\mathcal{L}_{\textbf{int}}=g\phi_A(x)\phi_B(x)\phi_C(x)$, where $g$ is a coupling constant and $\phi_i(x)$ is a scalar field evaluated at spacetime point $x$. This corresponds to the process $A\rightarrow B + C$. For a point interaction to happen, the scalar field $\phi_A(x)$ must be localized at the some spacetime point $x$. This must be the case because, in principle, we can trace the vertex of interaction by extrapolating the trajectory of decayed products (particles $B$ and $C$). Now, suppose a scalar particle $A$ is created at some localized point $q=q_0$ such that we can represent the field as a Gaussian. Under the evolution of some free-particle wave equation (Klein-Gordon or Schrödinger), the field is expected to spread out and not to be localized at some point. That is, point interaction will not happen and no particle should decay. However, if we invoke the GCM at the instant of particle $A$ creation, then the initial field will automatically collapse into one of the QTOA eigenfunctions with eigenvalue $\tau$. The unitary evolution, governed by the free wave equation, will naturally reduce the spread out field into a localized field at the interaction point $q=X$ and at the instant of interaction $t=\tau$. Notice that the time interval from the moment of particle creation to the instant of point interaction defines the lifetime of that particle. Furthermore, within the particle's lifetime, the field is evolving freely. Thus, the free quantum arrival time can be interpreted as a measure of lifetime.

In line with this, we now discuss the neutron lifetime anomaly in the framework of QTOA theory. From the above discussion, we can say that the neutron can only interact with the proton, electron, and antineutrino fields if the neutron's wave function is localized at the point of decay. Now, there is a small but noticeable difference in the lifetime of neutron based on two experimental methods \cite{Hirota:2020mrd,Wietfeldt:2011suo}. The first one is the so called beam experiment where the neutron's lifetime is measured to be $888.0$ seconds \cite{Mampe:1993an,Pichlmaier:2010zz,Steyerl:2012zz,Arzumanov:2015tea,Serebrov:2017jvb,Ezhov:2014tna}. The other one is the bottle experiment with measured lifetime of $879.4$ seconds \cite{Byrne:1996zz,Yue:2013qrc}. The main difference between the two methods is the way the initial states are prepared. Specifically, the beam and bottle experiments have different $k_0$ and possibly different $\sigma_0$ due to the temperature difference \cite{Heller2005}. Since the QTOA correction factor depends on the initial state, then the two preparations might be subject to different corrections. Proper estimation of the initial state parameters of the two methods must be obtained to verify if the observed anomaly is just a quantum effect and not a manifestation of new physics.

Alternatively, it is possible that the types of preparations correspond to two types of QTOA evolution shown in Fig.~\ref{fig:unitarycollapse}. The beam experiment is appropriately described by the left-right basis in \eqref{eq:pmbasis} which has been the focused of this study. On the other hand, the bottle experiment is best described by the even-odd basis in \eqref{eq:oddevenbasis}. The calculations of arrival time for states collapsing into the even-odd basis of the QTOA operator is not yet done and may not be straightforward because there is no classical counterpart to this set up. In addition, the initial state can collapse into the odd eigenfunctions which correspond to no interaction (or non-detection). For this specific intial state, the behavior of the arrival time as a function of initial state parameters may be completely different with the single-peak Gaussian wave packet. It is worth investigating how the no interaction part will affect the lifetime of an unstable particle or the expected detection time of an arriving particle. 

\section{Conclusion}

We updated the QTOA theory by imposing causality. In this updated theory, the quantum arrival time measurement proceed as follows. The initial state collapses right away into one of the eigenfunctions of the QTOA operator. The collapse is accompanied by the removal of the causality-violating component in the initial state. This is followed by unitary evolution of QTOA eigenfunction. The evolution proceeds from a spread out field to a localized distribution at the arrival point at the instant equal to the QTOA eigenvalue. This localization of the state at a specific instant of time is what we interpret as detection. As of the writing of this work, only the updated QTOA theory provides an intuitive physical picture of a time of arrival measurement. 

We also argue that the theory can be extended to describe the lifetime of an unstable particle. The preparation corresponds to the particle creation and the detection corresponds to the point interaction with the fields of the decay products. Such extension is valid because as long as the particle is freely propagating, standard quantum mechanics should provide a reasonable description. It is true that the vertex of interaction or decay point can happen anywhere and that the arrival description may not be fitting. But one can always count the number of decay events on a chosen point (interaction vertex) which effectively set the arrival point. Using this picture we also propose an explanation to the neutron's lifetime anomaly. That is, the observed discrepancy in the two measurements is a possible consequence of the quantum nature of time.

\section*{Acknowledgment}
This work was funded by the UP System Enhanced Creative Work and Research Grant (ECWRG-2021-2-12R).
NIS acknowledges the support of the Ateneo de Manila University through the Early Career Grant.


\bibliography{mybib}

\begin{thebibliography}{41}%
\makeatletter
\providecommand \@ifxundefined [1]{%
 \@ifx{#1\undefined}
}%
\providecommand \@ifnum [1]{%
 \ifnum #1\expandafter \@firstoftwo
 \else \expandafter \@secondoftwo
 \fi
}%
\providecommand \@ifx [1]{%
 \ifx #1\expandafter \@firstoftwo
 \else \expandafter \@secondoftwo
 \fi
}%
\providecommand \natexlab [1]{#1}%
\providecommand \enquote  [1]{``#1''}%
\providecommand \bibnamefont  [1]{#1}%
\providecommand \bibfnamefont [1]{#1}%
\providecommand \citenamefont [1]{#1}%
\providecommand \href@noop [0]{\@secondoftwo}%
\providecommand \href [0]{\begingroup \@sanitize@url \@href}%
\providecommand \@href[1]{\@@startlink{#1}\@@href}%
\providecommand \@@href[1]{\endgroup#1\@@endlink}%
\providecommand \@sanitize@url [0]{\catcode `\\12\catcode `\$12\catcode
  `\&12\catcode `\#12\catcode `\^12\catcode `\_12\catcode `\%12\relax}%
\providecommand \@@startlink[1]{}%
\providecommand \@@endlink[0]{}%
\providecommand \url  [0]{\begingroup\@sanitize@url \@url }%
\providecommand \@url [1]{\endgroup\@href {#1}{\urlprefix }}%
\providecommand \urlprefix  [0]{URL }%
\providecommand \Eprint [0]{\href }%
\providecommand \doibase [0]{https://doi.org/}%
\providecommand \selectlanguage [0]{\@gobble}%
\providecommand \bibinfo  [0]{\@secondoftwo}%
\providecommand \bibfield  [0]{\@secondoftwo}%
\providecommand \translation [1]{[#1]}%
\providecommand \BibitemOpen [0]{}%
\providecommand \bibitemStop [0]{}%
\providecommand \bibitemNoStop [0]{.\EOS\space}%
\providecommand \EOS [0]{\spacefactor3000\relax}%
\providecommand \BibitemShut  [1]{\csname bibitem#1\endcsname}%
\let\auto@bib@innerbib\@empty
\bibitem [{\citenamefont {Muga}\ \emph {et~al.}(2008)\citenamefont {Muga},
  \citenamefont {Mayato},\ and\ \citenamefont {Egusquiza}}]{Muga2008}%
  \BibitemOpen
  \bibfield  {author} {\bibinfo {author} {\bibfnamefont {J.~G.}\ \bibnamefont
  {Muga}}, \bibinfo {author} {\bibfnamefont {R.~S.}\ \bibnamefont {Mayato}},\
  and\ \bibinfo {author} {\bibfnamefont {I.~L.}\ \bibnamefont {Egusquiza}},\
  }\bibinfo {title} {Introduction},\ in\ \href
  {https://doi.org/10.1007/978-3-540-73473-4_1} {\emph {\bibinfo {booktitle}
  {Time in Quantum Mechanics}}},\ \bibinfo {editor} {edited by\ \bibinfo
  {editor} {\bibfnamefont {J.}~\bibnamefont {Muga}}, \bibinfo {editor}
  {\bibfnamefont {R.~S.}\ \bibnamefont {Mayato}},\ and\ \bibinfo {editor}
  {\bibfnamefont {{\'I}.}~\bibnamefont {Egusquiza}}}\ (\bibinfo  {publisher}
  {Springer Berlin Heidelberg},\ \bibinfo {address} {Berlin, Heidelberg},\
  \bibinfo {year} {2008})\ pp.\ \bibinfo {pages} {1--30}\BibitemShut {NoStop}%
\bibitem [{\citenamefont {Can\'ario}\ \emph {et~al.}(2021)\citenamefont
  {Can\'ario}, \citenamefont {Klaiber}, \citenamefont {Hatsagortsyan},\ and\
  \citenamefont {Keitel}}]{Reflection}%
  \BibitemOpen
  \bibfield  {author} {\bibinfo {author} {\bibfnamefont {D.~B.}\ \bibnamefont
  {Can\'ario}}, \bibinfo {author} {\bibfnamefont {M.}~\bibnamefont {Klaiber}},
  \bibinfo {author} {\bibfnamefont {K.~Z.}\ \bibnamefont {Hatsagortsyan}},\
  and\ \bibinfo {author} {\bibfnamefont {C.~H.}\ \bibnamefont {Keitel}},\
  }\bibfield  {title} {\bibinfo {title} {Role of reflections in the generation
  of a time delay in strong-field ionization},\ }\href
  {https://doi.org/10.1103/PhysRevA.104.033103} {\bibfield  {journal} {\bibinfo
   {journal} {Phys. Rev. A}\ }\textbf {\bibinfo {volume} {104}},\ \bibinfo
  {pages} {033103} (\bibinfo {year} {2021})}\BibitemShut {NoStop}%
\bibitem [{\citenamefont {Sainadh}\ \emph {et~al.}(2019)\citenamefont
  {Sainadh}, \citenamefont {Xu}, \citenamefont {Wang}, \citenamefont
  {Atia-Tul-Noor}, \citenamefont {Wallace}, \citenamefont {Douguet},
  \citenamefont {Bray}, \citenamefont {Ivanov}, \citenamefont {Bartschat},
  \citenamefont {Kheifets}, \citenamefont {Sang},\ and\ \citenamefont
  {Litvinyuk}}]{Htunneling}%
  \BibitemOpen
  \bibfield  {author} {\bibinfo {author} {\bibfnamefont {U.~S.}\ \bibnamefont
  {Sainadh}}, \bibinfo {author} {\bibfnamefont {H.}~\bibnamefont {Xu}},
  \bibinfo {author} {\bibfnamefont {X.}~\bibnamefont {Wang}}, \bibinfo {author}
  {\bibfnamefont {A.}~\bibnamefont {Atia-Tul-Noor}}, \bibinfo {author}
  {\bibfnamefont {W.~C.}\ \bibnamefont {Wallace}}, \bibinfo {author}
  {\bibfnamefont {N.}~\bibnamefont {Douguet}}, \bibinfo {author} {\bibfnamefont
  {A.}~\bibnamefont {Bray}}, \bibinfo {author} {\bibfnamefont {I.}~\bibnamefont
  {Ivanov}}, \bibinfo {author} {\bibfnamefont {K.}~\bibnamefont {Bartschat}},
  \bibinfo {author} {\bibfnamefont {A.}~\bibnamefont {Kheifets}}, \bibinfo
  {author} {\bibfnamefont {R.~T.}\ \bibnamefont {Sang}},\ and\ \bibinfo
  {author} {\bibfnamefont {I.~V.}\ \bibnamefont {Litvinyuk}},\ }\bibfield
  {title} {\bibinfo {title} {Attosecond angular streaking and tunnelling time
  in atomic hydrogen},\ }\href {https://doi.org/10.1038/s41586-019-1028-3}
  {\bibfield  {journal} {\bibinfo  {journal} {Nature}\ }\textbf {\bibinfo
  {volume} {568}},\ \bibinfo {pages} {75} (\bibinfo {year} {2019})}\BibitemShut
  {NoStop}%
\bibitem [{\citenamefont {Eckle}\ \emph {et~al.}(2008)\citenamefont {Eckle},
  \citenamefont {Pfeiffer}, \citenamefont {Cirelli}, \citenamefont {Staudte},
  \citenamefont {Dörner}, \citenamefont {Muller}, \citenamefont {Büttiker},\
  and\ \citenamefont {Keller}}]{Eckle}%
  \BibitemOpen
  \bibfield  {author} {\bibinfo {author} {\bibfnamefont {P.}~\bibnamefont
  {Eckle}}, \bibinfo {author} {\bibfnamefont {A.~N.}\ \bibnamefont {Pfeiffer}},
  \bibinfo {author} {\bibfnamefont {C.}~\bibnamefont {Cirelli}}, \bibinfo
  {author} {\bibfnamefont {A.}~\bibnamefont {Staudte}}, \bibinfo {author}
  {\bibfnamefont {R.}~\bibnamefont {Dörner}}, \bibinfo {author} {\bibfnamefont
  {H.~G.}\ \bibnamefont {Muller}}, \bibinfo {author} {\bibfnamefont
  {M.}~\bibnamefont {Büttiker}},\ and\ \bibinfo {author} {\bibfnamefont
  {U.}~\bibnamefont {Keller}},\ }\bibfield  {title} {\bibinfo {title}
  {Attosecond ionization and tunneling delay time measurements in helium},\
  }\href {https://doi.org/10.1126/science.1163439} {\bibfield  {journal}
  {\bibinfo  {journal} {Science}\ }\textbf {\bibinfo {volume} {322}},\ \bibinfo
  {pages} {1525} (\bibinfo {year} {2008})},\ \Eprint
  {https://arxiv.org/abs/https://www.science.org/doi/pdf/10.1126/science.1163439}
  {https://www.science.org/doi/pdf/10.1126/science.1163439} \BibitemShut
  {NoStop}%
\bibitem [{\citenamefont {Sombillo}\ and\ \citenamefont
  {Galapon}(2018)}]{Sombillo2018}%
  \BibitemOpen
  \bibfield  {author} {\bibinfo {author} {\bibfnamefont {D.~L.~B.}\
  \bibnamefont {Sombillo}}\ and\ \bibinfo {author} {\bibfnamefont {E.~A.}\
  \bibnamefont {Galapon}},\ }\bibfield  {title} {\bibinfo {title}
  {Barrier-traversal-time operator and the time-energy uncertainty relation},\
  }\href {https://doi.org/10.1103/PhysRevA.97.062127} {\bibfield  {journal}
  {\bibinfo  {journal} {Phys. Rev. A}\ }\textbf {\bibinfo {volume} {97}},\
  \bibinfo {pages} {062127} (\bibinfo {year} {2018})}\BibitemShut {NoStop}%
\bibitem [{\citenamefont {Pablico}\ and\ \citenamefont
  {Galapon}(2020)}]{PablicoTunnel}%
  \BibitemOpen
  \bibfield  {author} {\bibinfo {author} {\bibfnamefont {D.~A.~L.}\
  \bibnamefont {Pablico}}\ and\ \bibinfo {author} {\bibfnamefont {E.~A.}\
  \bibnamefont {Galapon}},\ }\bibfield  {title} {\bibinfo {title} {Quantum
  traversal time across a potential well},\ }\href
  {https://doi.org/10.1103/PhysRevA.101.022103} {\bibfield  {journal} {\bibinfo
   {journal} {Phys. Rev. A}\ }\textbf {\bibinfo {volume} {101}},\ \bibinfo
  {pages} {022103} (\bibinfo {year} {2020})}\BibitemShut {NoStop}%
\bibitem [{\citenamefont {Flores}\ and\ \citenamefont
  {Galapon}(2023)}]{Flores:2022pjn}%
  \BibitemOpen
  \bibfield  {author} {\bibinfo {author} {\bibfnamefont {P.~C.}\ \bibnamefont
  {Flores}}\ and\ \bibinfo {author} {\bibfnamefont {E.~A.}\ \bibnamefont
  {Galapon}},\ }\bibfield  {title} {\bibinfo {title} {{Instantaneous tunneling
  of relativistic massive spin-0 particles}},\ }\href
  {https://doi.org/10.1209/0295-5075/acad9a} {\bibfield  {journal} {\bibinfo
  {journal} {EPL}\ }\textbf {\bibinfo {volume} {141}},\ \bibinfo {pages}
  {10001} (\bibinfo {year} {2023})},\ \Eprint
  {https://arxiv.org/abs/2207.09040} {arXiv:2207.09040 [quant-ph]} \BibitemShut
  {NoStop}%
\bibitem [{\citenamefont {Das}\ and\ \citenamefont
  {D{\"u}rr}(2019)}]{DasArrival}%
  \BibitemOpen
  \bibfield  {author} {\bibinfo {author} {\bibfnamefont {S.}~\bibnamefont
  {Das}}\ and\ \bibinfo {author} {\bibfnamefont {D.}~\bibnamefont {D{\"u}rr}},\
  }\bibfield  {title} {\bibinfo {title} {Arrival time distributions of spin-1/2
  particles},\ }\href {https://doi.org/10.1038/s41598-018-38261-4} {\bibfield
  {journal} {\bibinfo  {journal} {Scientific Reports}\ }\textbf {\bibinfo
  {volume} {9}},\ \bibinfo {pages} {2242} (\bibinfo {year} {2019})}\BibitemShut
  {NoStop}%
\bibitem [{\citenamefont {Fadel}\ and\ \citenamefont
  {Maccone}(2021)}]{MacconePRA}%
  \BibitemOpen
  \bibfield  {author} {\bibinfo {author} {\bibfnamefont {M.}~\bibnamefont
  {Fadel}}\ and\ \bibinfo {author} {\bibfnamefont {L.}~\bibnamefont
  {Maccone}},\ }\bibfield  {title} {\bibinfo {title} {Time-energy uncertainty
  relation for quantum events},\ }\href
  {https://doi.org/10.1103/PhysRevA.104.L050204} {\bibfield  {journal}
  {\bibinfo  {journal} {Phys. Rev. A}\ }\textbf {\bibinfo {volume} {104}},\
  \bibinfo {pages} {L050204} (\bibinfo {year} {2021})}\BibitemShut {NoStop}%
\bibitem [{\citenamefont {Maccone}\ and\ \citenamefont
  {Sacha}(2020)}]{Maccone2020}%
  \BibitemOpen
  \bibfield  {author} {\bibinfo {author} {\bibfnamefont {L.}~\bibnamefont
  {Maccone}}\ and\ \bibinfo {author} {\bibfnamefont {K.}~\bibnamefont
  {Sacha}},\ }\bibfield  {title} {\bibinfo {title} {Quantum measurements of
  time},\ }\href {https://doi.org/10.1103/PhysRevLett.124.110402} {\bibfield
  {journal} {\bibinfo  {journal} {Phys. Rev. Lett.}\ }\textbf {\bibinfo
  {volume} {124}},\ \bibinfo {pages} {110402} (\bibinfo {year}
  {2020})}\BibitemShut {NoStop}%
\bibitem [{\citenamefont {Kazemi}\ and\ \citenamefont
  {Hosseinzadeh}(2023)}]{ExpTOA1}%
  \BibitemOpen
  \bibfield  {author} {\bibinfo {author} {\bibfnamefont {M.~J.}\ \bibnamefont
  {Kazemi}}\ and\ \bibinfo {author} {\bibfnamefont {V.}~\bibnamefont
  {Hosseinzadeh}},\ }\bibfield  {title} {\bibinfo {title} {Detection statistics
  in a double-double-slit experiment},\ }\href
  {https://doi.org/10.1103/PhysRevA.107.012223} {\bibfield  {journal} {\bibinfo
   {journal} {Phys. Rev. A}\ }\textbf {\bibinfo {volume} {107}},\ \bibinfo
  {pages} {012223} (\bibinfo {year} {2023})}\BibitemShut {NoStop}%
\bibitem [{\citenamefont {Flores}\ and\ \citenamefont
  {Galapon}(2019)}]{Flores:2018ken}%
  \BibitemOpen
  \bibfield  {author} {\bibinfo {author} {\bibfnamefont {P.~C.~M.}\
  \bibnamefont {Flores}}\ and\ \bibinfo {author} {\bibfnamefont {E.~A.}\
  \bibnamefont {Galapon}},\ }\bibfield  {title} {\bibinfo {title} {{Quantum
  free-fall motion and quantum violation of the weak equivalence principle}},\
  }\href {https://doi.org/10.1103/PhysRevA.99.042113} {\bibfield  {journal}
  {\bibinfo  {journal} {Phys. Rev. A}\ }\textbf {\bibinfo {volume} {99}},\
  \bibinfo {pages} {042113} (\bibinfo {year} {2019})},\ \Eprint
  {https://arxiv.org/abs/1808.02646} {arXiv:1808.02646 [quant-ph]} \BibitemShut
  {NoStop}%
\bibitem [{\citenamefont {Flores}\ \emph {et~al.}(2016)\citenamefont {Flores},
  \citenamefont {Caballar},\ and\ \citenamefont
  {Galapon}}]{PhysRevA.94.032123}%
  \BibitemOpen
  \bibfield  {author} {\bibinfo {author} {\bibfnamefont {P.~C.~M.}\
  \bibnamefont {Flores}}, \bibinfo {author} {\bibfnamefont {R.~C.~F.}\
  \bibnamefont {Caballar}},\ and\ \bibinfo {author} {\bibfnamefont {E.~A.}\
  \bibnamefont {Galapon}},\ }\bibfield  {title} {\bibinfo {title}
  {Synchronizing quantum and classical clocks made of quantum particles},\
  }\href {https://doi.org/10.1103/PhysRevA.94.032123} {\bibfield  {journal}
  {\bibinfo  {journal} {Phys. Rev. A}\ }\textbf {\bibinfo {volume} {94}},\
  \bibinfo {pages} {032123} (\bibinfo {year} {2016})}\BibitemShut {NoStop}%
\bibitem [{\citenamefont {Flores}\ and\ \citenamefont
  {Galapon}(2022)}]{Flores:2022vvk}%
  \BibitemOpen
  \bibfield  {author} {\bibinfo {author} {\bibfnamefont {P.~C.~M.}\
  \bibnamefont {Flores}}\ and\ \bibinfo {author} {\bibfnamefont {E.~A.}\
  \bibnamefont {Galapon}},\ }\bibfield  {title} {\bibinfo {title}
  {{Relativistic free-motion time-of-arrival operator for massive spin-0
  particles with positive energy}},\ }\href
  {https://doi.org/10.1103/PhysRevA.105.062208} {\bibfield  {journal} {\bibinfo
   {journal} {Phys. Rev. A}\ }\textbf {\bibinfo {volume} {105}},\ \bibinfo
  {pages} {062208} (\bibinfo {year} {2022})},\ \Eprint
  {https://arxiv.org/abs/2203.00898} {arXiv:2203.00898 [quant-ph]} \BibitemShut
  {NoStop}%
\bibitem [{\citenamefont {Pablico}\ and\ \citenamefont
  {Galapon}(2023)}]{Pablico:2022vjq}%
  \BibitemOpen
  \bibfield  {author} {\bibinfo {author} {\bibfnamefont {D.~A.~L.}\
  \bibnamefont {Pablico}}\ and\ \bibinfo {author} {\bibfnamefont {E.~A.}\
  \bibnamefont {Galapon}},\ }\bibfield  {title} {\bibinfo {title} {{Quantum
  corrections to the Weyl quantization of the classical time of arrival}},\
  }\href {https://doi.org/10.1140/epjp/s13360-023-03774-z} {\bibfield
  {journal} {\bibinfo  {journal} {Eur. Phys. J. Plus}\ }\textbf {\bibinfo
  {volume} {138}},\ \bibinfo {pages} {153} (\bibinfo {year} {2023})},\ \Eprint
  {https://arxiv.org/abs/2205.08694} {arXiv:2205.08694 [quant-ph]} \BibitemShut
  {NoStop}%
\bibitem [{\citenamefont {Galapon}(2009{\natexlab{a}})}]{Galapon2009}%
  \BibitemOpen
  \bibfield  {author} {\bibinfo {author} {\bibfnamefont {E.~A.}\ \bibnamefont
  {Galapon}},\ }\bibfield  {title} {\bibinfo {title} {Quantum wave-packet size
  effects on neutron time-of-flight spectroscopy},\ }\href
  {https://doi.org/10.1103/PhysRevA.80.030102} {\bibfield  {journal} {\bibinfo
  {journal} {Phys. Rev. A}\ }\textbf {\bibinfo {volume} {80}},\ \bibinfo
  {pages} {030102} (\bibinfo {year} {2009}{\natexlab{a}})}\BibitemShut
  {NoStop}%
\bibitem [{\citenamefont {Das}\ and\ \citenamefont
  {Nöth}(2021)}]{DasNoth2021}%
  \BibitemOpen
  \bibfield  {author} {\bibinfo {author} {\bibfnamefont {S.}~\bibnamefont
  {Das}}\ and\ \bibinfo {author} {\bibfnamefont {M.}~\bibnamefont {Nöth}},\
  }\bibfield  {title} {\bibinfo {title} {Times of arrival and gauge
  invariance},\ }\href {https://doi.org/10.1098/rspa.2021.0101} {\bibfield
  {journal} {\bibinfo  {journal} {Proc. R. Soc. A}\ }\textbf {\bibinfo {volume}
  {477}},\ \bibinfo {pages} {20210101} (\bibinfo {year} {2021})}\BibitemShut
  {NoStop}%
\bibitem [{\citenamefont {Galapon}(2009{\natexlab{b}})}]{Galapon2008.0278}%
  \BibitemOpen
  \bibfield  {author} {\bibinfo {author} {\bibfnamefont {E.~A.}\ \bibnamefont
  {Galapon}},\ }\bibfield  {title} {\bibinfo {title} {Theory of quantum arrival
  and spatial wave function collapse on the appearance of particle},\ }\href
  {https://doi.org/10.1098/rspa.2008.0278} {\bibfield  {journal} {\bibinfo
  {journal} {Proc. R. Soc. A}\ }\textbf {\bibinfo {volume} {465}},\ \bibinfo
  {pages} {71} (\bibinfo {year} {2009}{\natexlab{b}})}\BibitemShut {NoStop}%
\bibitem [{\citenamefont {Aharonov}\ and\ \citenamefont
  {Bohm}(1961)}]{AharonovBohm1961}%
  \BibitemOpen
  \bibfield  {author} {\bibinfo {author} {\bibfnamefont {Y.}~\bibnamefont
  {Aharonov}}\ and\ \bibinfo {author} {\bibfnamefont {D.}~\bibnamefont
  {Bohm}},\ }\bibfield  {title} {\bibinfo {title} {Time in the quantum theory
  and the uncertainty relation for time and energy},\ }\href
  {https://doi.org/10.1103/PhysRev.122.1649} {\bibfield  {journal} {\bibinfo
  {journal} {Phys. Rev.}\ }\textbf {\bibinfo {volume} {122}},\ \bibinfo {pages}
  {1649} (\bibinfo {year} {1961})}\BibitemShut {NoStop}%
\bibitem [{\citenamefont {Giannitrapani}(1997)}]{Giannitrapani1997}%
  \BibitemOpen
  \bibfield  {author} {\bibinfo {author} {\bibfnamefont {R.}~\bibnamefont
  {Giannitrapani}},\ }\bibfield  {title} {\bibinfo {title}
  {Positive-operator-valued time observable in quantum mechanics.},\ }\href
  {https://doi.org/10.1007/BF02435757} {\bibfield  {journal} {\bibinfo
  {journal} {Int. J. Theor. Phys.}\ }\textbf {\bibinfo {volume} {36}},\
  \bibinfo {pages} {1575} (\bibinfo {year} {1997})}\BibitemShut {NoStop}%
\bibitem [{\citenamefont {Galapon}(2004)}]{Galapon2004}%
  \BibitemOpen
  \bibfield  {author} {\bibinfo {author} {\bibfnamefont {E.~A.}\ \bibnamefont
  {Galapon}},\ }\bibfield  {title} {\bibinfo {title} {Shouldn’t there be an
  antithesis to quantization?},\ }\href {https://doi.org/10.1063/1.1767297}
  {\bibfield  {journal} {\bibinfo  {journal} {J. Math. Phys.}\ }\textbf
  {\bibinfo {volume} {45}},\ \bibinfo {pages} {3180} (\bibinfo {year}
  {2004})}\BibitemShut {NoStop}%
\bibitem [{\citenamefont {Muga}\ \emph {et~al.}(1998)\citenamefont {Muga},
  \citenamefont {Leavens},\ and\ \citenamefont {Palao}}]{MugaLeavensPalao1998}%
  \BibitemOpen
  \bibfield  {author} {\bibinfo {author} {\bibfnamefont {J.~G.}\ \bibnamefont
  {Muga}}, \bibinfo {author} {\bibfnamefont {C.~R.}\ \bibnamefont {Leavens}},\
  and\ \bibinfo {author} {\bibfnamefont {J.~P.}\ \bibnamefont {Palao}},\
  }\bibfield  {title} {\bibinfo {title} {Space-time properties of free-motion
  time-of-arrival eigenfunctions},\ }\href
  {https://doi.org/10.1103/PhysRevA.58.4336} {\bibfield  {journal} {\bibinfo
  {journal} {Phys. Rev. A}\ }\textbf {\bibinfo {volume} {58}},\ \bibinfo
  {pages} {4336} (\bibinfo {year} {1998})}\BibitemShut {NoStop}%
\bibitem [{\citenamefont {Sombillo}\ and\ \citenamefont
  {Galapon}(2016)}]{SOMBILLO2016261}%
  \BibitemOpen
  \bibfield  {author} {\bibinfo {author} {\bibfnamefont {D.~L.~B.}\
  \bibnamefont {Sombillo}}\ and\ \bibinfo {author} {\bibfnamefont {E.~A.}\
  \bibnamefont {Galapon}},\ }\bibfield  {title} {\bibinfo {title} {Particle
  detection and non-detection in a quantum time of arrival measurement},\
  }\href {https://doi.org/https://doi.org/10.1016/j.aop.2015.11.008} {\bibfield
   {journal} {\bibinfo  {journal} {Ann. Phys.}\ }\textbf {\bibinfo {volume}
  {364}},\ \bibinfo {pages} {261} (\bibinfo {year} {2016})}\BibitemShut
  {NoStop}%
\bibitem [{Foo()}]{Footnote1}%
  \BibitemOpen
  \href@noop {} {}\bibinfo {howpublished} {In our updated formulation, there is
  no need to invoke the confinement requirement in the original GCM. Unitary
  collapse of QTOA eigenfunction does not require confinement of the real
  position line.}\BibitemShut {Stop}%
\bibitem [{\citenamefont {Peskin}\ and\ \citenamefont
  {Schroeder}(1995)}]{Peskin}%
  \BibitemOpen
  \bibfield  {author} {\bibinfo {author} {\bibfnamefont {M.}~\bibnamefont
  {Peskin}}\ and\ \bibinfo {author} {\bibfnamefont {D.~V.}\ \bibnamefont
  {Schroeder}},\ }\href {https://doi.org/10.1201/9780429503559} {\emph
  {\bibinfo {title} {An Introduction To Quantum Field Theory}}}\ (\bibinfo
  {publisher} {Addison-Wesley Publishing Company},\ \bibinfo {year}
  {1995})\BibitemShut {NoStop}%
\bibitem [{\citenamefont {Duck}\ and\ \citenamefont
  {Sudarshan}(1998)}]{PauliSpin}%
  \BibitemOpen
  \bibfield  {author} {\bibinfo {author} {\bibfnamefont {I.}~\bibnamefont
  {Duck}}\ and\ \bibinfo {author} {\bibfnamefont {E.~C.~G.}\ \bibnamefont
  {Sudarshan}},\ }\href {https://doi.org/10.1142/3457} {\emph {\bibinfo {title}
  {Pauli and the Spin-Statistics Theorem}}}\ (\bibinfo  {publisher} {WORLD
  SCIENTIFIC},\ \bibinfo {year} {1998})\ \Eprint
  {https://arxiv.org/abs/https://www.worldscientific.com/doi/pdf/10.1142/3457}
  {https://www.worldscientific.com/doi/pdf/10.1142/3457} \BibitemShut {NoStop}%
\bibitem [{\citenamefont {Streater}\ and\ \citenamefont
  {Wightman}(2000)}]{PCT}%
  \BibitemOpen
  \bibfield  {author} {\bibinfo {author} {\bibfnamefont {R.}~\bibnamefont
  {Streater}}\ and\ \bibinfo {author} {\bibfnamefont {A.}~\bibnamefont
  {Wightman}},\ }\href
  {https://press.princeton.edu/books/paperback/9780691070629/pct-spin-and-statistics-and-all-that}
  {\emph {\bibinfo {title} {PCT, Spin and Statistics, and All that}}},\
  Princeton landmarks in mathematics and physics\ (\bibinfo  {publisher}
  {Princeton University Press},\ \bibinfo {year} {2000})\BibitemShut {NoStop}%
\bibitem [{\citenamefont {Eden}\ \emph {et~al.}(1966)\citenamefont {Eden},
  \citenamefont {Landshoff}, \citenamefont {Olive},\ and\ \citenamefont
  {Polkinghorne}}]{Eden:1966dnq}%
  \BibitemOpen
  \bibfield  {author} {\bibinfo {author} {\bibfnamefont {R.~J.}\ \bibnamefont
  {Eden}}, \bibinfo {author} {\bibfnamefont {P.~V.}\ \bibnamefont {Landshoff}},
  \bibinfo {author} {\bibfnamefont {D.~I.}\ \bibnamefont {Olive}},\ and\
  \bibinfo {author} {\bibfnamefont {J.~C.}\ \bibnamefont {Polkinghorne}},\
  }\href@noop {} {\emph {\bibinfo {title} {{The Analytic S-matrix}}}}\
  (\bibinfo  {publisher} {Cambridge Univ. Press},\ \bibinfo {address}
  {Cambridge},\ \bibinfo {year} {1966})\BibitemShut {NoStop}%
\bibitem [{\citenamefont {Newton}(1982)}]{Newton:1982qc}%
  \BibitemOpen
  \bibfield  {author} {\bibinfo {author} {\bibfnamefont {R.~G.}\ \bibnamefont
  {Newton}},\ }\href@noop {} {\emph {\bibinfo {title} {{Scathering Theory of
  Waves and Particles}}}}\ (\bibinfo  {publisher} {Springer Science+Business
  Media},\ \bibinfo {address} {New York},\ \bibinfo {year} {1982})\BibitemShut
  {NoStop}%
\bibitem [{\citenamefont {Sombillo}\ \emph {et~al.}(2021)\citenamefont
  {Sombillo}, \citenamefont {Ikeda}, \citenamefont {Sato},\ and\ \citenamefont
  {Hosaka}}]{Sombillo:2021ifs}%
  \BibitemOpen
  \bibfield  {author} {\bibinfo {author} {\bibfnamefont {D.~L.~B.}\
  \bibnamefont {Sombillo}}, \bibinfo {author} {\bibfnamefont {Y.}~\bibnamefont
  {Ikeda}}, \bibinfo {author} {\bibfnamefont {T.}~\bibnamefont {Sato}},\ and\
  \bibinfo {author} {\bibfnamefont {A.}~\bibnamefont {Hosaka}},\ }\bibfield
  {title} {\bibinfo {title} {{Classifying Near-Threshold Enhancement Using Deep
  Neural Network}},\ }\href {https://doi.org/10.1007/s00601-021-01642-z}
  {\bibfield  {journal} {\bibinfo  {journal} {Few Body Syst.}\ }\textbf
  {\bibinfo {volume} {62}},\ \bibinfo {pages} {52} (\bibinfo {year} {2021})},\
  \Eprint {https://arxiv.org/abs/2106.03453} {arXiv:2106.03453 [hep-ph]}
  \BibitemShut {NoStop}%
\bibitem [{\citenamefont {Hirota}\ \emph {et~al.}(2020)\citenamefont {Hirota}
  \emph {et~al.}}]{Hirota:2020mrd}%
  \BibitemOpen
  \bibfield  {author} {\bibinfo {author} {\bibfnamefont {K.}~\bibnamefont
  {Hirota}} \emph {et~al.},\ }\bibfield  {title} {\bibinfo {title} {{Neutron
  lifetime measurement with pulsed cold neutrons}},\ }\href
  {https://doi.org/10.1093/ptep/ptaa169} {\bibfield  {journal} {\bibinfo
  {journal} {PTEP}\ }\textbf {\bibinfo {volume} {2020}},\ \bibinfo {pages}
  {123C02} (\bibinfo {year} {2020})},\ \Eprint
  {https://arxiv.org/abs/2007.11293} {arXiv:2007.11293 [hep-ex]} \BibitemShut
  {NoStop}%
\bibitem [{\citenamefont {Wietfeldt}\ and\ \citenamefont
  {Greene}(2011)}]{Wietfeldt:2011suo}%
  \BibitemOpen
  \bibfield  {author} {\bibinfo {author} {\bibfnamefont {F.~E.}\ \bibnamefont
  {Wietfeldt}}\ and\ \bibinfo {author} {\bibfnamefont {G.~L.}\ \bibnamefont
  {Greene}},\ }\bibfield  {title} {\bibinfo {title} {{Colloquium: The neutron
  lifetime}},\ }\href {https://doi.org/10.1103/RevModPhys.83.1173} {\bibfield
  {journal} {\bibinfo  {journal} {Rev. Mod. Phys.}\ }\textbf {\bibinfo {volume}
  {83}},\ \bibinfo {pages} {1173} (\bibinfo {year} {2011})}\BibitemShut
  {NoStop}%
\bibitem [{\citenamefont {Mampe}\ \emph {et~al.}(1993)\citenamefont {Mampe},
  \citenamefont {Bondarenko}, \citenamefont {Morozov}, \citenamefont {Panin},\
  and\ \citenamefont {Fomin}}]{Mampe:1993an}%
  \BibitemOpen
  \bibfield  {author} {\bibinfo {author} {\bibfnamefont {W.}~\bibnamefont
  {Mampe}}, \bibinfo {author} {\bibfnamefont {L.~N.}\ \bibnamefont
  {Bondarenko}}, \bibinfo {author} {\bibfnamefont {V.~I.}\ \bibnamefont
  {Morozov}}, \bibinfo {author} {\bibfnamefont {Y.~N.}\ \bibnamefont {Panin}},\
  and\ \bibinfo {author} {\bibfnamefont {A.~I.}\ \bibnamefont {Fomin}},\
  }\bibfield  {title} {\bibinfo {title} {{Measuring neutron lifetime by storing
  ultracold neutrons and detecting inelastically scattered neutrons}},\
  }\href@noop {} {\bibfield  {journal} {\bibinfo  {journal} {JETP Lett.}\
  }\textbf {\bibinfo {volume} {57}},\ \bibinfo {pages} {82} (\bibinfo {year}
  {1993})}\BibitemShut {NoStop}%
\bibitem [{\citenamefont {Pichlmaier}\ \emph {et~al.}(2010)\citenamefont
  {Pichlmaier}, \citenamefont {Varlamov}, \citenamefont {Schreckenbach},\ and\
  \citenamefont {Geltenbort}}]{Pichlmaier:2010zz}%
  \BibitemOpen
  \bibfield  {author} {\bibinfo {author} {\bibfnamefont {A.}~\bibnamefont
  {Pichlmaier}}, \bibinfo {author} {\bibfnamefont {V.}~\bibnamefont
  {Varlamov}}, \bibinfo {author} {\bibfnamefont {K.}~\bibnamefont
  {Schreckenbach}},\ and\ \bibinfo {author} {\bibfnamefont {P.}~\bibnamefont
  {Geltenbort}},\ }\bibfield  {title} {\bibinfo {title} {{Neutron lifetime
  measurement with the UCN trap-in-trap MAMBO II}},\ }\href
  {https://doi.org/10.1016/j.physletb.2010.08.032} {\bibfield  {journal}
  {\bibinfo  {journal} {Phys. Lett. B}\ }\textbf {\bibinfo {volume} {693}},\
  \bibinfo {pages} {221} (\bibinfo {year} {2010})}\BibitemShut {NoStop}%
\bibitem [{\citenamefont {Steyerl}\ \emph {et~al.}(2012)\citenamefont
  {Steyerl}, \citenamefont {Pendlebury}, \citenamefont {Kaufman}, \citenamefont
  {Malik},\ and\ \citenamefont {Desai}}]{Steyerl:2012zz}%
  \BibitemOpen
  \bibfield  {author} {\bibinfo {author} {\bibfnamefont {A.}~\bibnamefont
  {Steyerl}}, \bibinfo {author} {\bibfnamefont {J.~M.}\ \bibnamefont
  {Pendlebury}}, \bibinfo {author} {\bibfnamefont {C.}~\bibnamefont {Kaufman}},
  \bibinfo {author} {\bibfnamefont {S.~S.}\ \bibnamefont {Malik}},\ and\
  \bibinfo {author} {\bibfnamefont {A.~M.}\ \bibnamefont {Desai}},\ }\bibfield
  {title} {\bibinfo {title} {{Quasielastic scattering in the interaction of
  ultracold neutrons with a liquid wall and application in a reanalysis of the
  Mambo I neutron-lifetime experiment}},\ }\href
  {https://doi.org/10.1103/PhysRevC.85.065503} {\bibfield  {journal} {\bibinfo
  {journal} {Phys. Rev. C}\ }\textbf {\bibinfo {volume} {85}},\ \bibinfo
  {pages} {065503} (\bibinfo {year} {2012})}\BibitemShut {NoStop}%
\bibitem [{\citenamefont {Arzumanov}\ \emph {et~al.}(2015)\citenamefont
  {Arzumanov}, \citenamefont {Bondarenko}, \citenamefont {Chernyavsky},
  \citenamefont {Geltenbort}, \citenamefont {Morozov}, \citenamefont
  {Nesvizhevsky}, \citenamefont {Panin},\ and\ \citenamefont
  {Strepetov}}]{Arzumanov:2015tea}%
  \BibitemOpen
  \bibfield  {author} {\bibinfo {author} {\bibfnamefont {S.}~\bibnamefont
  {Arzumanov}}, \bibinfo {author} {\bibfnamefont {L.}~\bibnamefont
  {Bondarenko}}, \bibinfo {author} {\bibfnamefont {S.}~\bibnamefont
  {Chernyavsky}}, \bibinfo {author} {\bibfnamefont {P.}~\bibnamefont
  {Geltenbort}}, \bibinfo {author} {\bibfnamefont {V.}~\bibnamefont {Morozov}},
  \bibinfo {author} {\bibfnamefont {V.~V.}\ \bibnamefont {Nesvizhevsky}},
  \bibinfo {author} {\bibfnamefont {Y.}~\bibnamefont {Panin}},\ and\ \bibinfo
  {author} {\bibfnamefont {A.}~\bibnamefont {Strepetov}},\ }\bibfield  {title}
  {\bibinfo {title} {{A measurement of the neutron lifetime using the method of
  storage of ultracold neutrons and detection of inelastically up-scattered
  neutrons}},\ }\href {https://doi.org/10.1016/j.physletb.2015.04.021}
  {\bibfield  {journal} {\bibinfo  {journal} {Phys. Lett. B}\ }\textbf
  {\bibinfo {volume} {745}},\ \bibinfo {pages} {79} (\bibinfo {year}
  {2015})}\BibitemShut {NoStop}%
\bibitem [{\citenamefont {Serebrov}\ \emph {et~al.}(2017)\citenamefont
  {Serebrov} \emph {et~al.}}]{Serebrov:2017jvb}%
  \BibitemOpen
  \bibfield  {author} {\bibinfo {author} {\bibfnamefont {A.~P.}\ \bibnamefont
  {Serebrov}} \emph {et~al.},\ }\bibfield  {title} {\bibinfo {title} {{New
  measurement of the neutron lifetime with a large gravitational trap}},\
  }\href {https://doi.org/10.1134/S0021364017220143} {\bibfield  {journal}
  {\bibinfo  {journal} {JETP Lett.}\ }\textbf {\bibinfo {volume} {106}},\
  \bibinfo {pages} {623} (\bibinfo {year} {2017})}\BibitemShut {NoStop}%
\bibitem [{\citenamefont {Ezhov}\ \emph {et~al.}(2018)\citenamefont {Ezhov}
  \emph {et~al.}}]{Ezhov:2014tna}%
  \BibitemOpen
  \bibfield  {author} {\bibinfo {author} {\bibfnamefont {V.~F.}\ \bibnamefont
  {Ezhov}} \emph {et~al.},\ }\bibfield  {title} {\bibinfo {title} {{Measurement
  of the neutron lifetime with ultra-cold neutrons stored in a
  magneto-gravitational trap}},\ }\href
  {https://doi.org/10.1134/S0021364018110024} {\bibfield  {journal} {\bibinfo
  {journal} {JETP Lett.}\ }\textbf {\bibinfo {volume} {107}},\ \bibinfo {pages}
  {671} (\bibinfo {year} {2018})},\ \Eprint {https://arxiv.org/abs/1412.7434}
  {arXiv:1412.7434 [nucl-ex]} \BibitemShut {NoStop}%
\bibitem [{\citenamefont {Byrne}\ and\ \citenamefont
  {Dawber}(1996)}]{Byrne:1996zz}%
  \BibitemOpen
  \bibfield  {author} {\bibinfo {author} {\bibfnamefont {J.}~\bibnamefont
  {Byrne}}\ and\ \bibinfo {author} {\bibfnamefont {P.~G.}\ \bibnamefont
  {Dawber}},\ }\bibfield  {title} {\bibinfo {title} {{A Revised Value for the
  Neutron Lifetime Measured Using a Penning Trap}},\ }\href
  {https://doi.org/10.1209/epl/i1996-00319-x} {\bibfield  {journal} {\bibinfo
  {journal} {EPL}\ }\textbf {\bibinfo {volume} {33}},\ \bibinfo {pages} {187}
  (\bibinfo {year} {1996})}\BibitemShut {NoStop}%
\bibitem [{\citenamefont {Yue}\ \emph {et~al.}(2013)\citenamefont {Yue},
  \citenamefont {Dewey}, \citenamefont {Gilliam}, \citenamefont {Greene},
  \citenamefont {Laptev}, \citenamefont {Nico}, \citenamefont {Snow},\ and\
  \citenamefont {Wietfeldt}}]{Yue:2013qrc}%
  \BibitemOpen
  \bibfield  {author} {\bibinfo {author} {\bibfnamefont {A.~T.}\ \bibnamefont
  {Yue}}, \bibinfo {author} {\bibfnamefont {M.~S.}\ \bibnamefont {Dewey}},
  \bibinfo {author} {\bibfnamefont {D.~M.}\ \bibnamefont {Gilliam}}, \bibinfo
  {author} {\bibfnamefont {G.~L.}\ \bibnamefont {Greene}}, \bibinfo {author}
  {\bibfnamefont {A.~B.}\ \bibnamefont {Laptev}}, \bibinfo {author}
  {\bibfnamefont {J.~S.}\ \bibnamefont {Nico}}, \bibinfo {author}
  {\bibfnamefont {W.~M.}\ \bibnamefont {Snow}},\ and\ \bibinfo {author}
  {\bibfnamefont {F.~E.}\ \bibnamefont {Wietfeldt}},\ }\bibfield  {title}
  {\bibinfo {title} {{Improved Determination of the Neutron Lifetime}},\ }\href
  {https://doi.org/10.1103/PhysRevLett.111.222501} {\bibfield  {journal}
  {\bibinfo  {journal} {Phys. Rev. Lett.}\ }\textbf {\bibinfo {volume} {111}},\
  \bibinfo {pages} {222501} (\bibinfo {year} {2013})},\ \Eprint
  {https://arxiv.org/abs/1309.2623} {arXiv:1309.2623 [nucl-ex]} \BibitemShut
  {NoStop}%
\bibitem [{\citenamefont {Heller}\ \emph {et~al.}(2005)\citenamefont {Heller},
  \citenamefont {Aidala}, \citenamefont {LeRoy}, \citenamefont {Bleszynski},
  \citenamefont {Kalben}, \citenamefont {Westervelt}, \citenamefont
  {Maranowski},\ and\ \citenamefont {Gossard}}]{Heller2005}%
  \BibitemOpen
  \bibfield  {author} {\bibinfo {author} {\bibfnamefont {E.~J.}\ \bibnamefont
  {Heller}}, \bibinfo {author} {\bibfnamefont {K.~E.}\ \bibnamefont {Aidala}},
  \bibinfo {author} {\bibfnamefont {B.~J.}\ \bibnamefont {LeRoy}}, \bibinfo
  {author} {\bibfnamefont {A.~C.}\ \bibnamefont {Bleszynski}}, \bibinfo
  {author} {\bibfnamefont {A.}~\bibnamefont {Kalben}}, \bibinfo {author}
  {\bibfnamefont {R.~M.}\ \bibnamefont {Westervelt}}, \bibinfo {author}
  {\bibfnamefont {K.~D.}\ \bibnamefont {Maranowski}},\ and\ \bibinfo {author}
  {\bibfnamefont {A.~C.}\ \bibnamefont {Gossard}},\ }\bibfield  {title}
  {\bibinfo {title} {Thermal averages in a quantum point contact with a single
  coherent wave packet},\ }\href
  {https://doi.org/https://doi.org/10.1021/nl0504585} {\bibfield  {journal}
  {\bibinfo  {journal} {Nano Lett.}\ }\textbf {\bibinfo {volume} {5}},\
  \bibinfo {pages} {1285} (\bibinfo {year} {2005})}\BibitemShut {NoStop}%
\end{thebibliography}%



\end{document}